\documentclass[fleqn,usenatbib]{mnras}

\usepackage{newtxtext,newtxmath}
\usepackage[T1]{fontenc}

\DeclareRobustCommand{\VAN}[3]{#2}
\let\VANthebibliography\thebibliography
\def\thebibliography{\DeclareRobustCommand{\VAN}[3]{##3}\VANthebibliography}

\usepackage{graphicx, subcaption}	
\usepackage{amsmath}	
\usepackage{soul}     
\usepackage[dvipsnames]{xcolor}
\usepackage{caption}
\usepackage{multicol}
\usepackage{relsize}  
\usepackage{tabularx, multirow}
\usepackage{fix-cm}
\usepackage{orcidlink}
\usepackage{lineno}

\usepackage{verbatim}
\makeatletter
\renewcommand{\verbatim@font}{\normalfont}
\makeatother

\newcommand{\normLz}{\mathcal{V}_{\phi}}
\newcommand{\snhd}{SNhd}
\newcommand{\narlt}{\! < \!}
\newcommand{\narsim}{\! \sim \!}

\title[{The central Splash in the Milky Way}]{The Splash population confined to the Milky Way bar region}
\author[A. Pandey and O. Gerhard]{
Aakash Pandey,\thanks{E-mail: 
aakashpandey2771998@gmail.com}\orcidlink{0009-0006-5677-7347}
Ortwin Gerhard,\thanks{E-mail: gerhard@mpe.mpg.de}\orcidlink{0000-0003-3333-0033}
\\
Max-Planck-Institute fur Extraterrestrische Physik, Gießenbachstraße, 85748 Garching, Germany
}

\date{Accepted XXX. Received YYY; in original form ZZZ}

\pubyear{2026}

\begin{document}

\label{firstpage}
\pagerange{\pageref{firstpage}--\pageref{lastpage}}
\maketitle

\begin{abstract}
    We investigate the chemo-dynamical properties of in-situ stars in the Galactic bar region (central $5$ kpc), specifically the inner Galactic counterpart of the solar-neighbourhood Splash population (in situ halo). We use a sample of APOGEE DR17 red giants with Gaia astrometry and AstroNN distances, and separate bulge/bar stellar populations dynamically using a normalised time-averaged bar-frame angular momentum, $\normLz$. 
    Chemo-dynamical maps in the $\normLz$-[Fe/H] plane reveal a pronounced extension of the bar population towards $\normLz<0$ (non-bar-following orbits) within $-1.0<{\rm [Fe/H]}<-0.3$, characterised by high eccentricities, large $z_{\rm max}$ and an extended spheroidal orbit density distribution, and accounting for $\sim 30\%$ of all stars at these metallicities. These $\normLz<0$ stars have [Mg/Fe] and [Al/Fe] abundances similar to those of stars on ${\rm x_1}$ orbits at the same [Fe/H], and they  match the solar-neighbourhood Splash in its metallicity and abundance ratio distributions. We therefore refer to them as the {\it central Splash}. 
    The presence of a significant central Splash underscores the role of the Gaia-Sausage-Enceladus merger in the early history of the Galaxy. 
    We find a similar, weaker extension in the $\normLz$-[Fe/H] maps for extremely metal-rich stars, which are confined to the bulge. At metallicities [Fe/H] $<0.3$, we find no additional classical bulge population in these maps beyond the central Splash and proto-Galactic populations -- the high-$\alpha$ bulge appears to have evolved from the early high-$\alpha$ disc. 
\end{abstract}


\section{Introduction}

Major progress has recently been made in identifying key stages in Galactic history, using stellar kinematics and stellar abundances from Gaia and spectroscopic surveys.  Studies of nearby stars have provided evidence for the primordial spin-up signalling the formation of the Galactic disc \citep{Belokurov2022Dawn, Rix2022poor}, and for the early merger of the Milky Way with the Gaia-Sausage-Enceladus (GSE) satellite  \citep[][]{Helmi2018, Belokurov2018}, which created both the anisotropic accreted inner Galactic stellar halo, as well as an in situ halo of kicked out disc stars, the Splash \citep[][]{Gallart2019, DiMatteo2019splash, Belokurov2020}.

The Galactic bar region, commonly taken to be inside $\sim5$ kpc, is the region inside corotation where the bar dominates the dynamics. This region contains about $60\%$ of the Milky Ways's total stellar mass \citep[$\sim3 \times10^{10}\,{\rm M_{\odot}}$,][]{Portail_2017, Khoperskov2025disc}, and the primary component is the barred boxy/peanut (b/p) bulge. The presence of an additional spheroidal dispersion-dominated classical bulge has not been excluded \citep{Barbuy2018, Queiroz2021milky}.

In this region the early disc has been transformed into the bar, and the in-situ and accreted halo are interspersed with the Galactic bulge. There is evidence for a spin-up at similar metallicity as in the nearby disc \citep{Rix2022poor, Pandey2025}, and there are indications that the in-situ halo (Splash) continues inwards into this region \citep{Liao2024insights} as is predicted by simulations \citep{Grand2020, Orkney_Laporte_2026}. The goal of this paper is to study the Splash population in the bar region and to disentangle it from the bulge and the bar.

N-body models compared with observations of nearby galaxies, including the Milky Way, suggest that b/p bulges arise from so-called secular evolution of the disc, whereby a bar first forms and then evolves into the b/p bulge \citep{Combes1990, Bureau2005, Shen2010}. On the other hand, a classical bulge forms rapidly by gravitational collapse or hierarchical merging of smaller objects and dissipative gas processes \citep{Steinmetz1995, Brooks+Ch2016}. Composite bulges are not unexpected \citep{Athanassoula2005} and have also been observed \citep{Erwin2015, Erwin2021, BlanaDiaz2018}. The composition of the Galactic bulge therefore provides crucial insights into the formation history of the Milky Way.

Large-scale surveys have revealed that stars in the Galactic bar region exhibit a clear bimodality in the [Fe/H]-[Mg/Fe]
plane: a metal-rich low-[Mg/Fe] component ([Fe/H] $\gtrsim-0.3$, [Mg/Fe] $\lesssim0.15$) and a metal-intermediate high-[Mg/Fe] component ($-1\lesssim{\rm [Fe/H]}\lesssim0$, [Mg/Fe] $\gtrsim0.15$) \citep[][]{Gonzalenz2011Alpha, Rojas2019bimodal, Lian2020milky}. At low metallicities ([Fe/H] $<-1$), stars are part of the inner stellar halo and the proto-Galactic population \citep{Belokurov2022Dawn, Myeong2022, Arentsen2024orbital}.

There is consensus that the strongly peanut-shaped low-[Mg/Fe] bulge \citep{Wegg2013mapping, Portail_2017Chem, Khoperskov2024III} was built from the early low-Mg (thin) disc \citep[][]{Barbuy2018, Shen2020}, however the origin of the high-[Mg/Fe] Galactic bulge is more debated.
The two bulge components show differences in their structural \citep[][]{Ness2012origin, Johnson2022bdbs, Khoperskov2024III}, and kinematic properties \citep[][]{Babusiaux2010, RojasArriagada2020many, Wylie2021a2a, Queiroz2021milky}. However, the large majority of high-[Mg/Fe] stars, those with [Fe/H] $\gtrsim-0.6$, show bar-like rotational properties similar to that of the low-[Mg/Fe] bulge \cite[][]{Ness2016ApogeeKinematics, Wylie2021a2a, Pandey2025}, and a fatter, shorter barred orbital density with a thicker b/p bulge \cite[][]{Portail_2017Chem, Khoperskov2024III, Pandey2025}, suggesting secular evolution from an $\alpha$-enhanced disc.

High-[Mg/Fe] stars with $-1\lesssim{\rm [Fe/H]}\lesssim-0.6$, however, have lower mean rotational velocities and higher velocity dispersions \citep{Queiroz2021milky, Pandey2025}, with oval-shaped and boxy orbital density distributions \citep{Khoperskov2024III, Pandey2025}. These dynamical properties have fuelled the ongoing debate over whether the high-[Mg/Fe] bulge has a completely in-situ origin through evolution from the high-$\alpha$ (thick) disc \citep[][]{DiMatteo2016discOrigin, Fragkoudi2018disc, Debattista2017fractionation, Debattista2019formation}, or whether an additional pressure-supported classical bulge component is required \citep[][]{Queiroz2021milky, Barbuy2023, Nepal2025}.

Historically, the old ages ($\gtrsim10$ Gyr) of bulge stars, particularly those with [Fe/H] $\lesssim-0.5$ \citep[][]{Clarkson2008, Brown2010, Gennaro2015, Renzini2018}, have provided a key argument for associating the high-$\alpha$ population with a classical bulge. Recent age analyses, however, have weakened this interpretation. Using colour-magnitude diagram fitting of Gaia DR3 data for stars in the Solar vicinity, \cite{FernandezAlvar2025} found that the kinematically selected thick disc is predominantly older than 10 Gyr \citep[see also][]{Gallart2024}, demonstrating that an old age alone does not necessarily imply a classical-bulge origin.

In this paper, we argue that the non bar-like dynamical properties of the metal-poor $(-1<{\rm [Fe/H]}-0.6)$ part of the high-[Mg/Fe] bulge can be explained with the inner Galaxy counterpart of the in-situ halo, `Splash' population identified in the solar neighbourhood (\snhd) \citep[][]{Bonaca2017, Haywood2018disguise, Gallart2019, DiMatteo2019splash, Belokurov2020}.

The Splash stars predominantly occupy the metallicity range $-1 < {\rm [Fe/H]} < -0.3$ \citep{Myeong2022, LeeAyeon2023} and their detailed chemical abundances are nearly identical to those of the Milky Way high-$\alpha$ disc stars in the same metallicity range \citep{Belokurov2022Dawn, Kisku2025}. The low rotational velocity tail of the high-$\alpha$ disc runs into the Splash \citep[][]{Gallart2019, DiMatteo2019splash, Belokurov2020}. These authors argued that the Splash stars were originally born in the high-$\alpha$ disc of the Milky Way and were subsequently scattered onto low angular momentum, high eccentricity orbits by the GSE merger event.

Cosmological simulations of Milky Way–like haloes that undergo GSE-like merger events show that Splash-like stars form a spheroidal component whose density peaks in the central regions of their host galaxies \citep{Grand2020, Orkney_Laporte_2026}. These simulations also predict that this event is accompanied by a massive centrally concentrated starburst, which has not been identified so far in the Milky Way. Combining with observational constraints from Gaia and legacy surveys, \citet{Orkney_Laporte_2026} argued that the GSE merger in the Milky Way was likely a minor merger, with a mass ratio of $\sim 1~:~7$.

In this work, we investigate the inner Galaxy counterpart of the \snhd\ in-situ halo `Splash’ population. We combine the precise abundances of APOGEE DR17 in-situ stars \citep{Abdurro2022apogeeDR17} with an orbit-dynamical analysis in a realistic bar/bulge potential \citep{Portail_2017} in Section~\ref{sec: 2}. We provide a chemo-dynamical characterisation of the Splash population confined to the Galactic bar region and link it to the \snhd\ Splash in Section~\ref{sec: 3}, discuss our results in Section~\ref{sec: 4}, and conclude in Section~\ref{sec: 5}.

\section{Data: in-situ stars from APOGEE DR17}\label{sec: 2}
We use stars from the APOGEE spectroscopic survey\footnote{Apache Point Obs. Galactic Evolution Experiment, \cite{Majewski2016apogee}} data release DR17 \citep{Abdurro2022apogeeDR17}. The APOGEE survey was designed to obtain high-resolution ($\sim 22,\,500$) near-infrared ($1.5-1.7,\mu$m) spectra of stars across all major components of the Milky Way, allowing APOGEE to make observations even in highly dust-extincted regions and making it particularly well-suited for studying the inner Galaxy. In addition to line-of-sight velocities, APOGEE provides precise abundance measurements for up to $\sim 20$ chemical elements, covering a wide range of nucleosynthetic processes -- the aluminium-to-iron abundance ratios ([Al/Fe]) are key diagnostics for distinguishing between in-situ and accreted stellar populations in the Milky Way \citep{Hawkins2015chemical, Belokurov2022Dawn}.

The main survey targets were randomly selected for observation from a high-quality subset of the Two Micron All Sky Survey \citep[2MASS;][]{Skrutskie2006two}. For details on the APOGEE targeting strategies, see \cite{Zasowski2013ApogeeTarget, Zasowski2017ApogeeTarget}, \cite{Beaton2021final}, and \cite{Santana2021final}. Stellar parameters and chemical abundances for APOGEE stars are derived using the APOGEE Stellar Parameters and Chemical Abundances Pipeline \citep[ASPCAP;][]{Holtzman2015, Holtzman2018Apogee,Garcia_Perez2016}.

Chemical abundances and line-of-sight velocities from the APOGEE survey are supplemented by astrometry from Gaia EDR3 \citep{GaiaCollab2021EDR3} and spectro-photometric distances from the AstroNN value-added catalogue for APOGEE \citep{Leung2019astronn, Leung2023measurement}, resulting in full 6D phase-space information for individual stars.

We use the 6D phase-space information to compute the Galactocentric coordinates $(X,\,Y,\,Z)$, with cylindrical radius $R = \sqrt{X^2 + Y^2}$. We adopt the inertial frame Galactocentric cylindrical velocities $V_{\phi}$ (azimuthal) and $V_R$ (radial), where $V_{\phi}$ is defined as positive in the direction of Galactic rotation. For the coordinate transformation, we assume the Solar position to be $(X,\,Y,\,Z) = (-R_{\odot},\,0,\,Z_{\odot})$, with $R_{\odot}= 8.2$ kpc \citep{BlandHawthorn2016, Gravity2021, Leung2023measurement} and $Z_{\odot}= 20.8$ pc \citep{Bennett2019vertical}, and its Cartesian velocities $\mathbf{v}_{\odot} = (U_{\odot},\,V_0 + V_{\odot},\,W_{\odot}) = (11.1,\,248.54,\,7.25)\,\mathrm{km\,s^{-1}}$, where $U_{\odot}$ and $W_{\odot}$ are taken from \citet{Schoenrich2010local}, and $V_0 + V_{\odot}$ is from \citet{Reid2020sagittarius} based on the proper motion of Sgr~A.

\subsection{Base and bar samples} \label{sec: 21}
Our base sample consists of in-situ stars with reliable chemical abundances and precise distances. We first select stars from the APOGEE main sample by requiring \texttt{EXTRATARG == 0}. To ensure high-quality stellar parameters, we retain stars with a signal-to-noise ratio ${\rm S/N > 50}$ and with the 23rd bit of \texttt{ASPCAPFLAG} set to zero. To obtain a red giant star sample, we apply the following criteria: $\log g < 2$ (with uncertainty $< 0.2$) and effective temperature $T_{\rm eff} > 3200$ K. To exclude stars with potentially spurious abundance measurements, we require the uncertainties in [Fe/H], [Mg/Fe], and [Al/Fe] to be $<0.2$ dex. We also require the fractional uncertainties in the AstroNN distances to be $<20\%$.

From this cleaned sample, we identify in-situ stars following \cite{Belokurov2022Dawn}: ${\rm [Fe/H]} > -0.4$ \texttt{OR} ${\rm [Al/Fe]} > -0.075$. Further, \cite{Belokurov2020} found that Splash stars in the \snhd\ do not extend beyond $R \sim 15$ kpc; therefore, we exclude stars with $R > 15$ kpc. Together, these selection criteria define our {\it base sample} of 64,686 stars. To focus on stars confined to the bar region, we apply an additional cut on the cylindrical apocentre distance: $R_{\rm apo} \leq 5$ kpc (see Section~\ref{sec: 22}). This defines our {\it bar sample}, which retains approximately $22\%$ of the base sample. We note that approximately $97\%$ of the stars in our bar sample have fractional distance uncertainties of $<10\%$ and the uncertainties in [X/Fe] are $\lesssim0.03$ dex. Moreover, the above chemical selection to identify in-situ stars excludes approximately $7\%$ of stars from the base sample and $4\%$ from the bar sample.

\subsection{Galactic potential and orbits integration} \label{sec: 22}
For the dynamical analysis we integrate orbits of all APOGEE stars in our base sample in a gravitational potential from the suite of Made-to-Measure dynamical bulge-bar models developed by \citet{Portail_2017}. These models were adapted to fit star count density and kinematic data from the VVV, UKIDSS, 2MASS, BRAVA, OGLE, and ARGOS surveys. In the models, the distance from the Sun to the Galactic centre is taken to be $R_{\odot} = 8.2$ kpc, and the bar angle is $\alpha_{\odot} = 28^{\circ}$, based on the distribution of red clump stars from \citet{Wegg2013mapping, Wegg_2015}.

We adopt a potential model with a mass-to-red clump star number ratio of $M_*/N_{\rm RC}=1000\,M_{\odot}/$ RC star, a central in-plane mass $M_{\rm c} = 2 \times 10^9\,M_{\odot}$, and a pattern speed $\Omega_{\rm b} = 37.5\,{\rm km\,s^{-1}\,kpc^{-1}}$ as our standard case. This model provides a good fit to both the bulge proper motions \citep{Clarke2019virac} and the inner Galaxy gas distribution \citep{Li2022gas}. Moreover, \cite{Sormani_2022} provide an analytic approximation of the model density, which is available in the \texttt{AGAMA} library \citep{Vasiliev_2019}.

We also integrate orbits in an alternative potential model from \cite{Portail_2017} to confirm that our conclusions do not depend on a specific model (see Section~\ref{sec: 33}). This alternative model differs in the shape of the potential, the mass-to–red clump star number ratio is $M_*/N_{\rm RC}=1100\,M_{\odot}/$ RC star, the central in-plane mass is half that of the standard model, and the bar pattern speed is $\Omega_{\rm b} = 35\,{\rm km\,s^{-1}\,kpc^{-1}}$.

To integrate the orbit of a star, we treat the observed phase-space coordinates as initial conditions, transform them from the heliocentric frame to the rotating bar-frame, and integrate the orbit using the drift–kick–drift leapfrog algorithm implemented in the NMAGIC code \citep{DeLorenzi2007nmagic}, for a total time 2 Gyr saving orbit points every 1 Myr. 

\begin{figure*}
    \centering
    \includegraphics[]{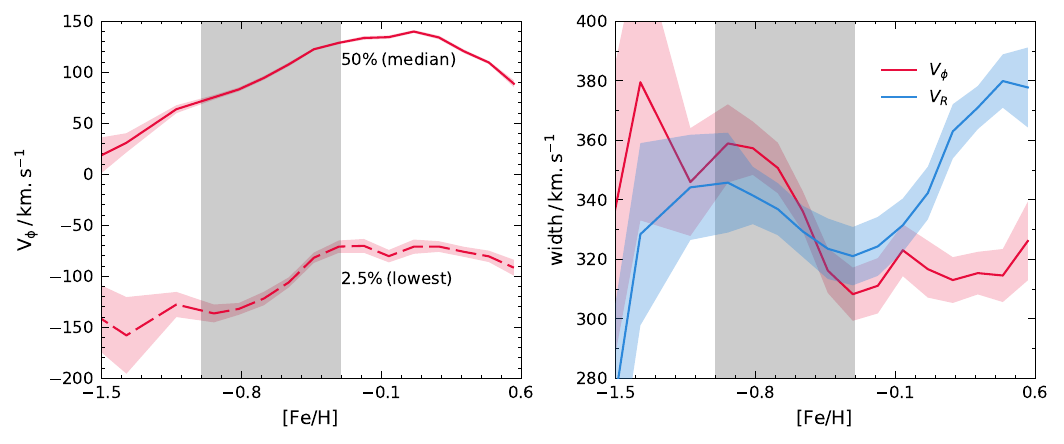}
\caption{Metallicity dependence of the median and lowest tail of the Galactocentric rotational velocity distribution (left), and the width $(\equiv 97.5\%-2.5\% \sim4.3\times \,{\rm dispersion})$ of the rotational and radial velocity distributions (right), for all stars in our bar sample. Velocity percentiles are computed using a boxcar kernel of width $0.125$ dex. The band widths indicate the bootstrapped uncertainties. The grey stripes mark the metallicity range of the \snhd\ Splash stars.}
    \label{fig: fig1}
\end{figure*}

For each orbit, we define the apocentre and pericentre of the cylindrical radial distance, $R_{\rm apo}$ and $R_{\rm peri}$, as the 90th and 10th percentiles of the time series of the cylindrical radius, respectively. Similarly, we define the apocentre of the spherical radial distance, $r_{\rm apo}$, and the maximum vertical excursion along the minor axis of the bar, $z_{\rm max}$, as the 90th percentile of the corresponding time series for $r$ and the absolute vertical coordinate. Using the cylindrical $R_{\rm apo}$ and $R_{\rm peri}$ we define the orbital eccentricity as $(R_{\rm apo} - R_{\rm peri}) / (R_{\rm apo} + R_{\rm peri})$.

To extend the orbital analysis, we compute the bar-frame Cartesian orbital frequencies ($f_x,\,f_y,\,f_z$) as well as the cylindrical radial and azimuthal frequencies ($f_R,\,f_{\phi}$) for all stars in our bar sample using \texttt{naif}, a Python package developed by \cite{BeraldoESilva2023} based on the NAFF algorithm (Numerical Analysis of Fundamental Frequencies; \citealp{Laskar1992, ValluriMerritt1998}). Following \cite{Papaphilippou1996, Papaphilippou1998} and \cite{BeraldoESilva2023}, we use the complex time series of each coordinate as input to extract the dominant frequency. A key advantage of this approach is that it automatically determines the sign of the cylindrical azimuthal frequency $f_{\phi}$ \citep[see][]{BeraldoESilva2023}, corresponding to the direction of the dominant rotational motion (prograde or retrograde)

Using these frequencies, we identify stars associated with the ${\rm x_1}$ family tree (the main bar-supporting orbits), the retrograde ${\rm x_4}$ orbits, and other orbit families. Following \cite{Pandey2025}, the ${\rm x_1}$ orbits are defined by $f_x/f_y = 1 \pm 0.1$ \texttt{AND} $f_x/f_z \leq 0.7$ \texttt{AND} $f_R/f_\phi = 2 \pm 0.1$, while the ${\rm x_4}$ orbits are identified by $f_x/f_y = 1 \pm 0.1$ \texttt{AND} $f_x/f_z > 0.7$ \texttt{AND} $f_{\phi} < 0$.

\section{The Splash population in the bar region}\label{sec: 3}

\subsection{Signature of Splash stars using observed kinematics}\label{sec: 31}
Splash stars identified in \snhd\ studies are characterised by low, and often retrograde, rotational velocities together with a large radial velocity dispersion \citep[][]{Gallart2019, DiMatteo2019splash, Belokurov2020, Chandra2024}. They are predominantly found in the metallicity range $-1<{\rm [Fe/H]<-0.3}$ \citep[][]{Myeong2022, LeeAyeon2023} which we refer to  in this work as the {\it Splash metallicity range} or {\it SplashMR} in short. 

To search for initial signatures of Splash stars in the bar region, we first examine differential changes with metallicity in the Galactocentric rotational and radial velocity distributions (RotVD and RadVD) of stars in our bar sample. We note that in the bar region, angular momentum is not conserved and negative rotational velocities can also occur for overall prograde orbits.

The left panel of Fig.~\ref{fig: fig1} shows the median and the lowest tail (2.5th percentile) of the RotVD of stars in our bar sample as a function of metallicity. Across the metallicity range $-0.3 \lesssim {\rm [Fe/H]} \lesssim 0.6$, the lowest tail is retrograde $(\sim -70 \, {\rm km\,s^{-1}})$ and remains nearly flat. The presence of retrograde rotating stars at such metallicities in the inner region has been reported in earlier studies \citep[e.g.][]{Queiroz2021milky}. As the metallicity drops below ${\rm [Fe/H]}\sim-0.3$, the lowest tail rapidly shifts to more retrograde values. The sharp drop in the lowest tail is accompanied by a steady decrease in the median rotational velocity, showing that the overall distribution shifts to lower rotational velocities.

The right panel of Fig.~\ref{fig: fig1} shows the metallicity dependence of the widths of the RotVD and RadVD, which we measure as the difference of the 97.5th and 2.5th percentiles ($\sim4.3~\times~\sigma$). As metallicity decreases from ${\rm [Fe/H]}\sim0.6$ to $\sim-0.3$, the width of the RadVD decreases, reflecting that the horizontal elongation of the barred density distribution decreases with decreasing metallicity \citep[][]{Portail_2017Chem, Khoperskov2024III}. At [Fe/H] $\sim -0.3$, the width of the RadVD turns around and increases in the SplashMR, indicating a surge in a radially hot population towards lower metallicities. This trend is also accompanied by an increase in the width of the RotVD. Below [Fe/H] $\sim-1$, the width of the RadVD again decreases, consistent with recent findings that metal poor stars become more centrally concentrated at lower metallicities \citep[][]{Rix2022poor}. 

The shift in the RotVD of our bar sample stars towards lower velocities in the SplashMR $-1 < {\rm [Fe/H]} < -0.3$, accompanied by the turnaround and simultaneous increase in the width of the RadVD, are reminiscent of the Splash population in the \snhd. These findings are also consistent with results from \cite{Liao2024insights} who analysed Gaia DR3 stars with low-resolution XP spectra in the $V_{\phi}-{\rm [Fe/H]_{XP}}$ plane across Galactocentric radii, from the bar region to $R\sim14\,{\rm kpc}$. They found a persistent extension of the $V_{\phi}$ distribution towards negative values in the metallicity range $-1 < {\rm [Fe/H]_{XP}} < -0.4$. Even though characterising the negative $V_\phi$ stars as part of the Splash population in the bar region is uncertain due to disc/halo interlopers and the influence of the non-axisymmetric potential, the radial consistency of this feature suggests a common origin across the entire radial range, such as the GSE merger.

\begin{figure*}
    \centering
    \includegraphics[]{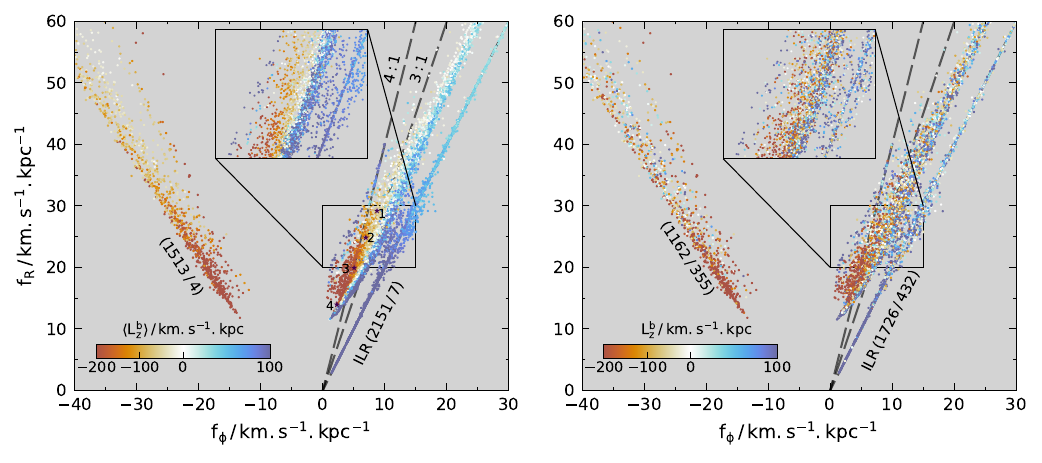}
    \caption{Scatter plot of all stars in our bar sample in the cylindrical azimuthal $f_{\phi}$ versus radial $f_R$ frequency plane, colour coded by the time-averaged bar-frame rotational angular momenta of their orbits $\langle L_z^{\rm b} \rangle$ (left), and by their present angular momenta $L_z^{\rm b}$ (right). For stars with $f_{\phi}<0$ and those near the inner Lindblad resonance $({\rm ILR},\,f_R:f_{\phi}=2:1)$, numbers in brackets indicate the counts of stars with (expected$\,/\,$opposite) sign of angular momentum for the respective orbit family. Insets show zoomed-in views of the dense region indicated. The $3:1$ and $4:1$ resonances are indicated by dashed black lines. Orbits marked with labels $1-4$ are typical of those with $\normLz<0$ but $f_{\phi}>0$, and are illustrated in Fig.~\ref{fig: figA1}.}
    \label{fig: fig2}
\end{figure*}

\subsection{Time-averaged bar-frame angular momentum} \label{sec: 32}
Splash stars are thought to originate predominantly from the oldest part of the high-$\alpha$ (thick) disc and to have been strongly heated by the GSE merger \citep{DiMatteo2019splash, Gallart2019, Belokurov2020, Orkney_Laporte_2026}. When the Galactic bar forms later-on \citep{Sanders2024bar, Haywood2024}, these stars would participate as a pre-existing dynamically hot population during the bar and buckling instabilities. Such populations respond to but do not drive the forming bar, and they are expected to contribute less to the resulting barred density and dynamics than the colder disc stars that cause the bar instability \citep{DiMatteo2016discOrigin, Perez2017stellar, Debattista2017fractionation}. Thus, also at much later times the Splash stars are expected to be dynamically distinct from the main underlying barred stellar populations evolved from the rotating disc.

However, the dynamical separation between these different stellar populations in the bar region is blurred by the rotating bar. In particular, the rotational angular momentum of a star along its orbit -- which is an approximate integral of motion in the \snhd\ -- is not conserved in the bar region due to the non-axisymmetric potential \citep{LyndenBellKalnajs1972, BT2008}. Consequently, even stars on the main bar-supporting (${\rm x_1}$) orbits can temporarily exhibit retrograde angular momentum \citep[e.g.][]{Valluri2016unified}. Thus simply identifying stars with retrograde rotational velocities in the bar region does not necessarily imply that they belong to the Splash population or, more generally, to the inner stellar halo.

$N$-body models of barred galaxies have shown overall positive mean streaming rotation velocities for bar stars in the bar-frame \citep[][]{Sellwood_1987, Qin_2015}, implying that bar stars have positive average bar-frame rotational angular momentum along their orbits. $N$-body models have further shown that the main bar-supporting ${\rm x_1}$ orbits have large net (time-averaged) rotational angular momentum in the bar-frame, and that other bar-supporting orbits also possess smaller positive net angular momentum \citep[][]{Valluri2016unified, Petersen2019}.

Therefore, as a means of distinguishing between dynamically different stellar populations, we investigate their time-averaged bar-frame rotational angular momentum $\langle L_z^{\rm b} \rangle$. For each star in our bar sample, we define $\langle L_z^{\rm b} \rangle$ as
\begin{equation}
	\label{eq: eq1}
	\langle L_z^{\rm b}\rangle = \frac{1}{N} \mathlarger{\sum}_i\left(R_i\,V_{\phi,\,i}-R_i^2\,\Omega_{\rm b}\right) \equiv \frac{1}{N} \mathlarger{\sum}_i \left( R_i\,V_{\phi,\,i}^{\rm b}\right)
\end{equation}
\noindent where $R_i$, $V_{\phi,\,i}$, and $V_{\phi,\,i}^{\rm b}$ are the Galactocentric radius, inertial and bar-frame rotational velocities at the $i$-th time step along its orbit, integrated for a total number of $N$ equally spaced time steps in the gravitational potential (see Section~\ref{sec: 22}). $\Omega_{\rm b}$ is the assumed bar pattern speed, and $V_{\phi,\,i}$ and $V_{\phi,\,i}^{\rm b}$ are related as $V_{\phi,\,i}=V_{\phi,\,i}^{\rm b}+R_i\,\Omega_{\rm b}$. The present-day bar-frame rotational angular momentum for each star is defined using its observed phase-space coordinates: $L_z^{\rm b}=R\,V_{\phi}-R^2\,\Omega_{\rm b}$. 

We first investigate how well the time-averaged $\langle L_z^{\rm b} \rangle$ corresponds to groups of similar orbits, characterizing each orbit by its orbital frequencies (see \ref{sec: 22}), particularly by its cylindrical azimuthal frequency $f_{\phi}$ and radial frequency $f_R$ associated with the in-plane motion. For quasi-periodic orbits, the fundamental frequencies can be used as integrals of motion to classify the orbits \citep[e.g.,][]{BT2008}. The leading frequencies computed in Section~\ref{sec: 22} using \texttt{naif} are linear combinations of the fundamental frequencies \citep{BinneySpergel1982, Papaphilippou1996, Papaphilippou1998}, and they likewise classify the orbits. This should still be approximately true for weakly chaotic orbits confined by and approximately following regular orbits, but would fail for strongly chaotic orbits.

Fig.~\ref{fig: fig2} shows the distribution of stars in the $f_{\phi}$-$f_R$ plane. The scattered points are colour coded by the time-averaged $\langle L_z^{\rm b} \rangle$ (left panel) and, for reference, by the observed $L_z^{\rm b}$ of the orbit (right panel). In both panels, the distribution is clearly separated into positive and negative $f_{\phi}$ regimes. We note that most stars with $f_{\phi} < 0$ belong to the retrograde ${\rm x_4}$ orbit family (see Section~\ref{sec: 35}). Stars with $f_{\phi} > 0$ are divided into two groups: those aligned along the $f_R : f_{\phi} = 2:1$ line (inner Lindblad resonance, ILR, the main bar-supporting orbits) and a cloud of near-resonant orbits located above this line \citep{Voglis_2007, Valluri2016unified}.

Interestingly, some orbits with $f_{\phi}>0$ have $\langle L_z^{\rm b} \rangle < 0$, whereas all orbits with $f_{\phi}<0$ also have $\langle L_z^{\rm b} \rangle < 0$. Indeed, approximately $13\%$ stars in our bar sample show this one-sided sign mismatch, while the remaining $87\%$ have matching signs \citep[see also][]{Prudil2025}. Inspection of individual orbits shows that these mismatched cases arise because the bar-frame rotational velocity is strongly positive near pericentre but negative near apocentre. Their orbital morphology evolves from round, three-dimensional orbits in the bulge to planar, multi-foil orbits in the outer bar. Fig.~\ref{fig: figA1} shows four typical orbits selected from Fig.~\ref{fig: fig2}.

\begin{figure}
	\centering
	\includegraphics[]{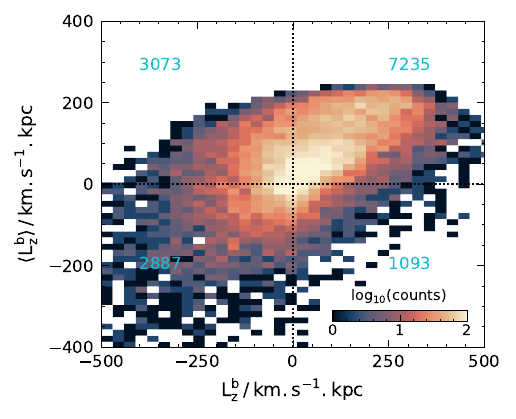}
	\caption{Star count distribution for our bar sample in the plane of time-averaged $\langle L_z^{\rm b}\rangle$ versus observed $L_z^{\rm b}$ bar-frame angular momentum. Dotted lines mark zero values and numbers indicate the counts in each quadrant.}
	\label{fig: fig3}
\end{figure}

The left panel of Fig.~\ref{fig: fig2} shows that there is almost no intermixing of positive and negative $\langle L_z^{\rm b} \rangle$ values in the orbit frequency distribution. Nearly all stars near the ILR -- the main bar-supporting orbits -- exhibit large positive values. For the near-resonant orbits, at a fixed $f_R$, $\langle L_z^{\rm b} \rangle$ gradually decreases from large positive to negative values as $f_{\phi}$ decreases, resulting in minimal intermixing. Further, all stars with $f_{\phi} < 0$ have $\langle L_z^{\rm b} \rangle < 0$. This suggests that the time-averaged bar-frame rotational angular momentum $\langle L_z^{\rm b} \rangle$ shares aspects of an approximate integral of motion in the barred potential\footnote{\citet{QinShen2021} earlier proposed $\langle L_z^{\rm b} \rangle\,/\,\sigma_{L_z^{\rm b}}$ as an approximate integral, where $\sigma_{L_z^{\rm b}}$ denotes the dispersion of $L_z^{\rm b}$ along the orbit.}; however we do not pursue this issue in more detail here, and in fact will use a scaled version of $\langle L_z^{\rm b} \rangle$ later-on in the paper.

On the other hand, the right column shows strong intermixing of positive and negative $L_z^{\rm b}$ values. For stars with $f_{\phi}<0$, approximately $24\%$ having $L_z^{\rm b}>0$. A similar intermixing of positive and negative $L_z^{\rm b}$ values is seen among stars near the ILR, with about $20\%$ having $L_z^{\rm b}<0$. This intermixing is even more apparent for the near-resonant orbits located above the inner Lindblad resonance.

The cause for the intermixing of the signs and magnitudes of $L_z^{\rm b}$ values seen in the right panel of Fig.~\ref{fig: fig2} is that the rotational angular momentum varies along the orbits. As illustrated in Fig.~\ref{fig: fig3}, the resulting star count distribution in the $L_z^{\rm b}$–$\langle L_z^{\rm b} \rangle$ plane shows only a weak correlation between $L_z^{\rm b}$ and $\langle L_z^{\rm b} \rangle$. Only for $70\%$ of stars the signs of $L_z^{\rm b}$ and $\langle L_z^{\rm b} \rangle$ agree and, notably, among stars with $L_z^{\rm b} < 0$, the majority have $\langle L_z^{\rm b} \rangle > 0$. 

While $\langle L_z^{\rm b} \rangle$ is therefore well-suited to characterize orbits in the bar region, equation~(\ref{eq: eq1}) shows that it explicitly involves the Galactocentric radius $R$ along the orbit, so that its numerical value depends strongly on the orbit's radial extent; see Fig.~\ref{fig: figA2}. Consequently, trends with respect to $\langle L_z^{\rm b} \rangle$ inherit an implicit dependence on the average Galactocentric radii of the observed sample stars. To mitigate this, we define a normalised time-averaged bar-frame rotational angular momentum as 
\begin{equation}
	\label{eq: eq2}
    \normLz \equiv \langle L_z^{\rm b} \rangle \big/ {\langle R \rangle} = \mathlarger{\sum}_i \, R_i\left(V_{\phi,\,i}-R_i\,\Omega_{\rm b}\right) \big/ {\mathlarger{\sum}_i R_i} 
\end{equation}
\noindent where $\langle R \rangle$ denotes the time-average of the Galactocentric radius along an orbit and the remaining symbols have the same meaning as in equation~\ref{eq: eq1}. Thus $\normLz$ is a radius-weighted average rotational velocity in the bar-frame, and keeps the same sign as $\langle L_z^{\rm b} \rangle$.

As illustrated in the right panel of Fig.~\ref{fig: figA2}, the normalised $\normLz$ is largely independent of radius although, overall, negative values occur more frequently in the bulge ($\lesssim 2.5$ kpc). Therefore, in the following we use $\normLz$ to characterize orbits, e.g., for Splash stars, and to define stellar subsamples in the bar region to be analysed further.

\subsection{Chemo-dynamical maps in the \texorpdfstring{$\normLz$}{angledVphi}-metallicity plane}\label{sec: 33}
\begin{figure*}
    \centering
    \includegraphics[]{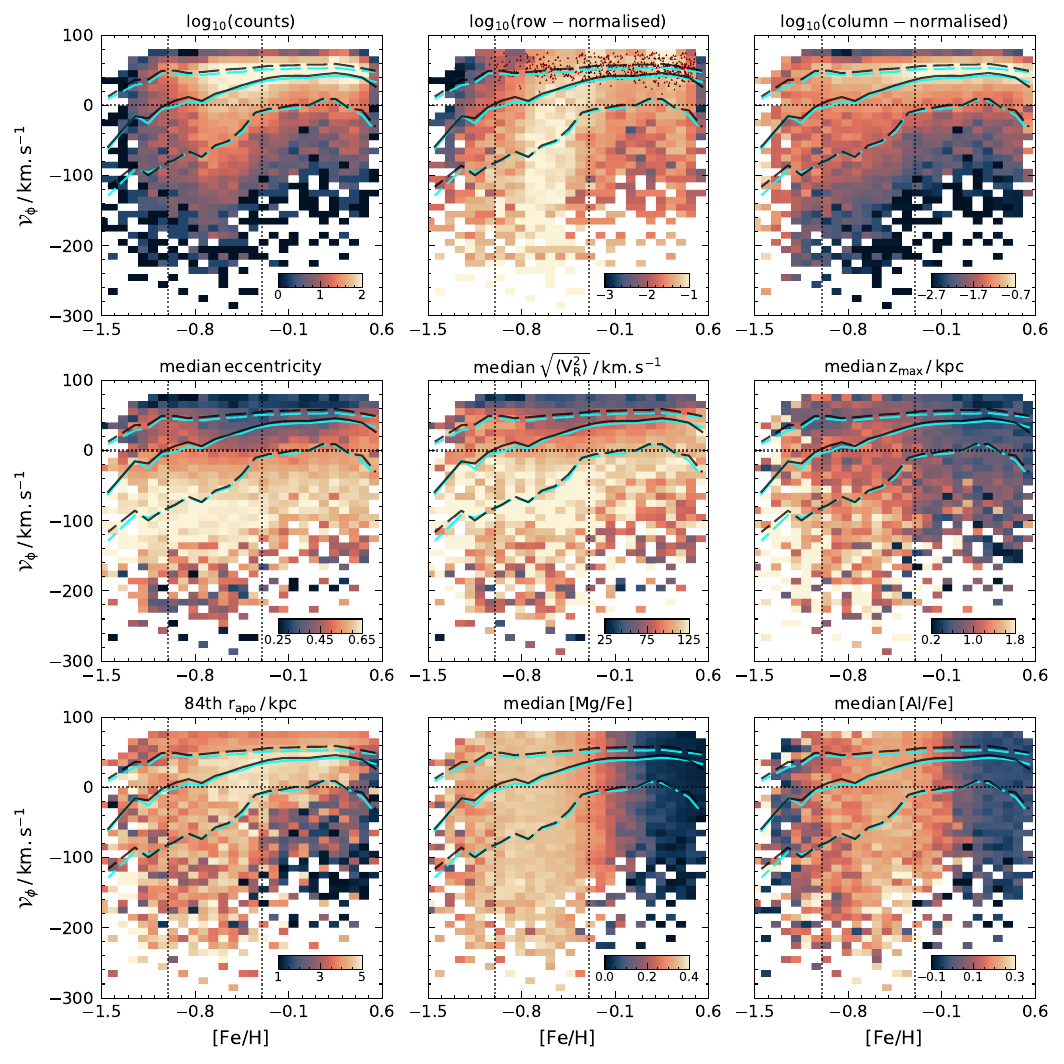}
    \caption{Top row: Star count distribution in the plane of normalised time-averaged bar-frame rotational angular momentum $\normLz$ versus metallicity (left), and the corresponding row- (central) and column-normalised (right) distributions for all stars in our bar sample. Maroon points in the central panel show stars from the ${\rm x_1}$ orbit family, only a random quarter is displayed for clarity. Middle and bottom rows: Dynamical and chemical properties across the [Fe/H]-$\normLz$ plane, as indicated in the panel titles, Corresponding values per pixel are shown by the respective colour bars. In all panels, the solid black line represents the running median of the entire $\normLz$ distribution with [Fe/H], and the upper and lower dashed black lines mark the 16th and 84th percentiles of the $\normLz$ distribution. The cyan curves show the same statistics, with $\normLz$ computed using an alternate potential model (see Section~\ref{sec: 22}). The vertical dotted black lines mark the SplashMR, and the horizontal dotted black line marks $\normLz = 0$.}
    \label{fig: fig4}
\end{figure*}
Fig.~\ref{fig: fig4} shows the dynamical and chemical properties of stars in our bar sample in the $\normLz$-[Fe/H] plane. The top row presents the counts distribution along with the corresponding row- and column-normalised distributions. Maroon scattered points in the row-normalised distribution represent stars from the ${\rm x_1}$ orbit family, with high positive $\normLz$ values. The middle and bottom rows show dynamical and chemical properties across this plane as indicated in the panel titles. The running median and the $16$th and $84$th percentile values of the entire $\normLz$ distribution vs [Fe/H] are shown in all panels as solid and dashed black lines, respectively. The cyan lines show the same statistics computed using an alternative potential model which differs in the shape of the potential and has a lower pattern speed (see Section~\ref{sec: 22}) -- the agreement between the black and cyan lines demonstrates that the $\normLz$-distributions are consistent between these potential models.

\paragraph*{Counts distribution} 
The counts distribution exhibits a nearly metallicity-independent, prominent peak at $\normLz\!\sim\!50\,{\rm km\,s^{-1}}$ (top-left). The distribution also shows extensions towards and into large negative $\normLz$ in two separate metallicity intervals: a dominant $\normLz < 0$ overdensity in the SplashMR ($-1<{\rm [Fe/H]}<-0.3$) with highest density in $-0.8<{\rm [Fe/H]}<-0.4$), and a sub-dominant overdensity at [Fe/H] $>0.25$. Stars with [Fe/H] $<-1$ are distributed more symmetrically around $\normLz\sim0$. 

The row-normalised distribution highlights the predominance of stars with $\normLz <0$ in the SplashMR (top-central). The column-normalised distribution (top-right) moreover reveals a vertical stretching towards large negative values of $\normLz$, which is clearly reflected in the sharp drop of the 16th percentile as metallicity decreases below [Fe/H] $\sim-0.3$ and the accompanied more gradual decline in the median. This stretching towards low $\normLz$ is less pronounced in the $\normLz <0$ overdensity at high metallicities. 

\paragraph*{Dynamical maps}
The distribution of the median orbital eccentricity in the $\normLz$-metallicity plane (middle-left) mainly varies with $\normLz$. The median eccentricity progressively increases as $\normLz$ decreases from the highest values to $\sim-100\,{\rm km\,s^{-1}}$, where orbital eccentricities are highest, and then declines at lower values. This decline at the lowest $\normLz$ values is driven by the retrograde ${\rm x_4}$ orbits. 

The middle-central panel shows the distribution of the median $\sqrt{\langle V_R^2 \rangle}$, computed by averaging the square of radial velocities along the orbit of each star. As expected, the distribution closely mirrors the median eccentricity distribution (middle-left), showing that stars with $-100<\normLz\,/\,{\rm km\,s^{-1}}<0$ in the SplashMR exhibit the highest radial velocity dispersion.

The distribution of the median values of the maximum orbital height $z_{\rm max}$ (middle-right) shows an overall increase with decreasing metallicity: for [Fe/H] $>-0.3$, the median $\rm z_{\rm max}$ is small ($<0.5$ kpc), then increases slowly with decreasing metallicity while remaining nearly independent of $\normLz$. In contrast, within and below the SplashMR, the median $z_{\rm max}$ rises gradually with decreasing $\normLz$, reaching up to $\sim 1.5$ kpc at the lowest values.   

The distribution of 84th percentile values for the spherical apocentre distance $r_{\rm apo}$ (median of upper 32\%, lower-left) shows that orbits for stars with $\normLz > 0$ extend to the outer bar, except for the most metal-rich ($\gtrsim0.4$) stars which are confined to the bulge ($<2.5$ kpc). Recently, \cite{Khoperskov2025disc} found using APOGEE DR17 data that the most metal rich stars are trapped in the inner $1$-$2$ kpc region \citep[see also][]{Rix2024EMRknot, Khoperskov2024III}. Stars with [Fe/H] $> -0.3$ and $\normLz < 0$ are also confined to the bulge.  

In contrast, orbits for stars in the SplashMR reach the outer bar for both positive and negative values of $\normLz$; those with positive values are slightly more extended. The abrupt change in the radial extent of orbits at [Fe/H] $\sim-0.3$ separates the two overdensities of stars with $\normLz < 0$ seen in the counts distribution (top row, left column). For [Fe/H] $<-1$, the orbits are slightly less radially extended than those in the SplashMR \citep[see also][]{Rix2022poor, Khoperskov2024III}. 

Thus, the dynamical maps reveal that the overdensity of stars with $\normLz<0$ in the SplashMR ($-1 < {\rm [Fe/H]} < -0.3$) is characterised by radially hot orbits with large vertical excursions, consistent with their dynamical connection to the \snhd\ Splash stars. The most metal-rich ([Fe/H] $\gtrsim 0.4$) stars are confined to the central $\sim 2.5$ kpc region. This property is also shared by stars with $\normLz < 0$ at metallicities as low as [Fe/H] $\sim-0.3$.

\paragraph*{Chemical maps}
The central and right panels in the bottom row of Fig.~\ref{fig: fig4} show the median values of the [Mg/Fe] and [Al/Fe] abundance ratios in the $\normLz$-metallicity plane. Both abundance ratios depend on metallicity as expected. For super-solar metallicity stars, the median [Mg/Fe] and [Al/Fe] are $\narsim 0$, from where they steeply increase towards lower metallicities, reaching $\narsim 0.3$ and\ $\!\sim0.2$ at [Fe/H] $\narsim-0.3$. Below [Fe/H] $\narsim-0.3$ the high-$\alpha$ population dominates, and the median [Mg/Fe] remain nearly unchanged down to [Fe/H] $\narsim-1$, decreasing only slightly at even lower metallicities. On the other hand, the median [Al/Fe] show a steep decrease below [Fe/H] $\narsim-0.7$ towards the lowest metallicities \citep[see also][]{Rojas2019bimodal, Belokurov2022Dawn}. 

However, the medians of both the abundance ratios show almost no dependence on $\normLz$ at a given [Fe/H]. Consequently, in the SplashMR $-1 \narlt {\rm [Fe/H]} \narlt -0.3$, stars with $\normLz <0$ are chemically analogous to those with positive $\normLz$. To investigate this further, Fig.~\ref{fig: figA3} shows that the very slight increase observed for the median [Mg/Fe] from $\narsim0.33$ at $\normLz \narsim75\,{\rm km\,s^{-1}}$ to $\narsim0.36$ at $\narsim-180\,{\rm km\,s^{-1}}$ remains within $\leq 1\sigma$ range of the distribution.

\begin{figure*}
    \centering
    \includegraphics[]{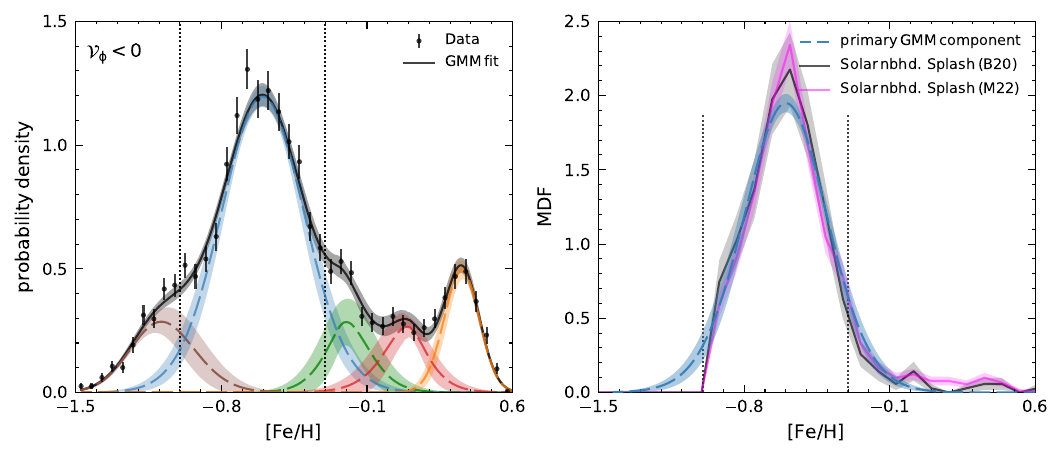}
    \caption{Left: MDF of stars with $\normLz < 0$ (black points), and its five-component Gaussian fit. The solid black line and the coloured dashed lines show the overall fit and the individual components. Right:  Comparison of the MDF of the primary Gaussian component (weight $\sim65\%$) with that for \snhd\ Splash stars. Cyan and magenta curves represent the \snhd\ Splash samples collectively selected from our base sample (see Section~\ref{sec: 21}) following \protect\cite{Belokurov2020} ($V_{\phi} < 0$ and [Fe/H] $>-1$) and \protect\cite{Myeong2022} (eccentricity $>0.6$ and [Fe/H] $>-1$) and requiring $R_{\rm apo}>5$ kpc. The dotted vertical lines in both columns mark [Fe/H] $=-1$ and $-0.3$.}
    \label{fig: fig5}
\end{figure*}

\begin{figure*}
    \centering
    \includegraphics[]{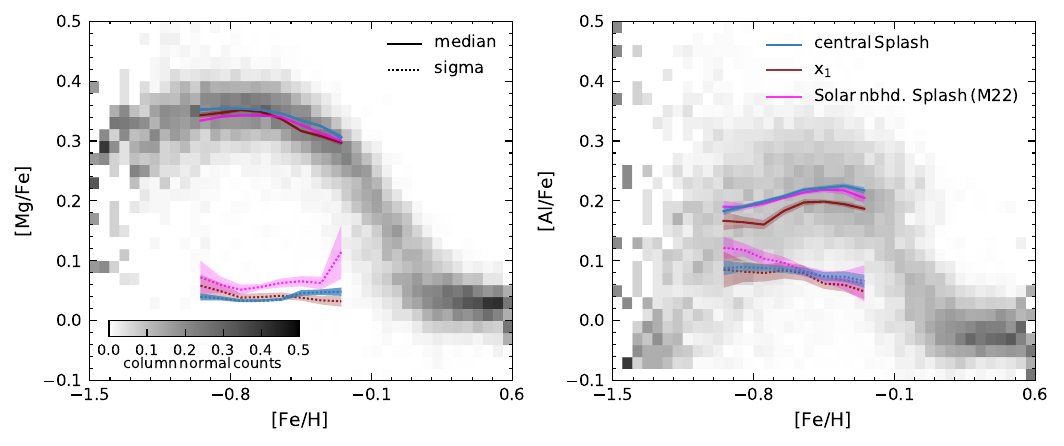}
\caption{The grey-scale shows the column normalised distribution of stars in our bar sample in the [Fe/H]-[X/Fe] plane. Solid and dotted lines indicate the median and sigma of the [X/Fe] distribution for the different stellar samples, with colours corresponding to the samples as indicated (see text for details). Here, sigma ($\sigma_{\rm dst}$) is defined as $(84{\rm th}-16{\rm th}) / 2$ percentiles width of the distribution in each metallicity bin. The band widths indicate the bootstrapped uncertainties.}
    \label{fig: fig6}
\end{figure*}

\paragraph*{Overall} the chemo-dynamical maps for our bar sample stars in the $\normLz$-metallicity plane reveal two main results. First, despite the fact that in the SplashMR ($-1 < {\rm [Fe/H]} < -0.3$) we see stars with quite different dynamical properties - those on positive $\normLz$ orbits, with fairly low eccentricities and low $z_{\rm max}$, and those with very negative $\normLz$, high orbital eccentricities and large $z_{\rm max}$ -- the  [Mg/Fe] and [Al/Fe] abundances remain nearly the same across the entire $\normLz$ range. This clearly suggests a common formation history, similar to that for the thick (high-$\alpha$) disc and Splash stars in the \snhd. 

Second, the most metal-rich stars (${\rm [Fe/H]} \gtrsim0.4$) also extend to quite negative $\normLz$, but are centrally confined to the inner bulge ($\sim2.5$ kpc). These stars are likely connected to the recently discovered central `knot' stars \citep{Rix2024EMRknot, Khoperskov2025disc}. In addition, a small subset of stars with $\normLz < 0$ and [Fe/H] $\gtrsim-0.3$ is also centrally confined to the inner bulge -- representing only $15\%$ of all stars in this metallicity range.

\subsection{The metallicity distribution and detailed chemical abundances: central Splash} \label{sec: 34}

\begin{figure*}
	\centering
	\includegraphics[]{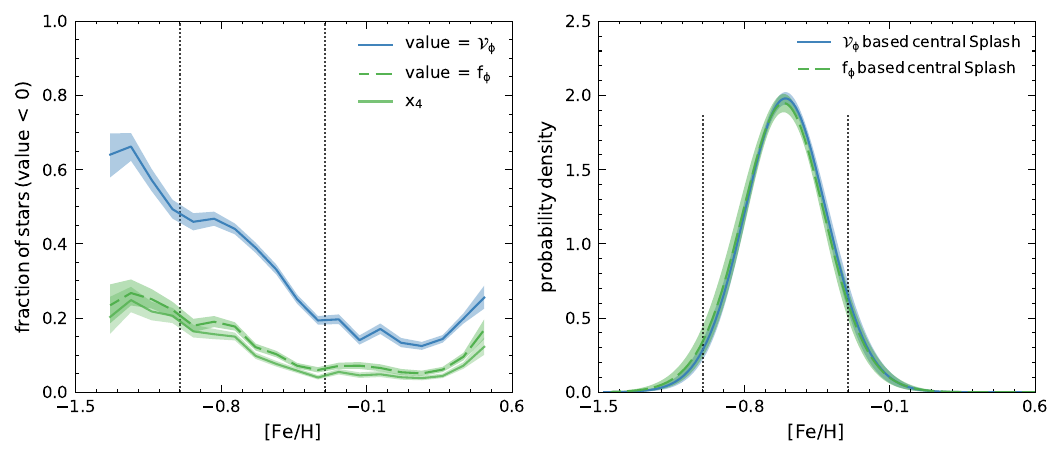}
	\caption{Left: Fraction of stars with $\normLz < 0$ (blue) and $f_{\phi} < 0$ (dashed green) versus metallicity. The solid green curve shows the fraction of stars linked to the retrograde ${\rm x_4}$ family. The bins at the lowest and highest metallicities are omitted because they are dominated by noise. Right: Primary Gaussian component from a five-component fit to the MDF of all stars in our bar sample with $f_{\phi} < 0$ (dashed green). The blue curve is the same as the primary component shown in Fig.~\ref{fig: fig5}.}
	\label{fig: fig7}
\end{figure*}

Here we further investigate the connection between bar region stars with $\normLz < 0$ in the metallicity range $-1 < {\rm [Fe/H]} < -0.3$, and Splash stars in the \snhd, by comparing their metallicity distribution functions (MDFs) as well as the distributions of [Mg/Fe] and [Al/Fe] abundance ratios.

The left panel of Fig.~\ref{fig: fig5} shows the MDF of stars with $\normLz < 0$ in our bar sample, along with a five-component Gaussian fit. The distribution exhibits a dominant peak at [Fe/H] $\sim-0.6$ within the SplashMR ($-1 < {\rm [Fe/H]} < -0.3$), along with sub-dominant peaks above and below this range. Notably, the dominant primary Gaussian component (with weight $\sim65\%$) accounts for nearly all stars in the SplashMR. 

The sub-dominant peaks at [Fe/H] $\sim0.35$ and $\sim0.1$ correspond to low-[Mg/Fe] ($\leq0.1$) stars, while the peak at [Fe/H] $\sim-0.2$ is associated predominantly with high-[Mg/Fe] ($\sim0.25$) stars. Finally, stars below [Fe/H] $\sim-1$ are represented by a Gaussian component peaking at [Fe/H] $\sim-1.1$, which predominantly belong to the proto-galactic population \citep{Belokurov2022Dawn, Rix2022poor, Myeong2022, Pandey2025}. Only the primary peak and the peak at [Fe/H] $\sim0.35$ can be clearly regarded as separate components.

The right column of Fig.~\ref{fig: fig5} compares the primary Gaussian component from the left column with the MDF of Splash stars in the \snhd. The \snhd\ Splash stars are collectively selected from our base sample (see Section~\ref{sec: 21}) following \cite{Belokurov2020}: [Fe/H] $>-1$ and $V_{\phi}<0$, and \cite{Myeong2022}: [Fe/H] $>-1$, eccentricity $>0.6$, and requiring $R_{\rm  apo}>5$ kpc. The comparison reveals a remarkably good agreement between the MDF of the primary component and that of the \snhd\ Splash stars, suggesting that both populations are closely interconnected.

Fig.~\ref{fig: fig6} extends the chemical comparison to the [Mg/Fe] and [Al/Fe] abundance ratio distributions. The grey-scale shows the column normalised distribution of all stars in our bar sample, and the solid and dotted lines indicate the running median and $1\sigma_{\rm dst}$ width for bar region stars with $\normLz < 0$ (blue), stars associated with the ${\rm x_1}$ bar orbit family (maroon), and \snhd\ Splash stars (magenta), over the common metallicity range $-1 < {\rm [Fe/H]} < -0.3$. The \snhd\ Splash sample is the same eccentricity based sample as used in Fig.~\ref{fig: fig5} \citep[following][]{Myeong2022}. For both abundance ratios, distribution medians and widths for the bar region stars with $\normLz < 0$ are in remarkably good agreement with those for the \snhd\ Splash stars. This further confirms that the two stellar populations are closely linked.

Moreover, Fig.~\ref{fig: fig6} shows that the [Mg/Fe] and [Al/Fe] distribution moments for stars with $\normLz < 0$ in the SplashMR are in good agreement with those of stars associated with the ${\rm x_1}$ orbit family in the same metallicity range, within $\lesssim 0.5\sigma_{\rm dst}$. This close agreement in detailed chemical abundances suggests a similar initial formation history for the $\normLz<0$-stellar population and the bar population in the SplashMR. The bar population is believed to originate in the early high-$\alpha$ disc which later evolved into the bar and b/p-bulge \citep{DiMatteo2016discOrigin, Debattista2017fractionation, Fragkoudi2018disc, Pandey2025}. Therefore the similar abundance ratios suggest that also the $\normLz<0$ -stars evolved from the early high-$\alpha$ disc.

Overall, the agreement between the MDF of bar region stars with $\normLz < 0$ and $-1 < {\rm [Fe/H]} < -0.3$, and the MDF of \snhd\ Splash stars, implies that both populations are closely linked, further reinforced by their matching detailed abundances. Additionally, the similar matching with the abundance ratios of the $\rm x_1$ stars in the SplashMR links the origin of these  $\normLz < 0$ stars to the early high-$\alpha$ disc. Although bar/bulge instabilities may also contribute to $\normLz < 0$ stars in the SplashMR, in Section~\ref{sec: 37} we argue that this contribution remains small. Therefore, we conclude that bar region stars with $\normLz < 0$ and $-1 < {\rm [Fe/H]} < -0.3$ predominantly represent the inner-Galaxy counterpart of the \snhd\ Splash population, which we hereafter refer to as the {\it central Splash}.

\subsection{Retrograde central Splash stars} \label{sec: 35}
In the previous section, we identified central Splash stars by their negative values of $\normLz$. Fig.~\ref{fig: fig2} shows that orbits with $\normLz<0$ separate into two broad branches. The first branch is dominated by the widely studied retrograde ${\rm x_4}$ orbit family which have $f_{\phi}<0$. The second branch comprises orbits with mildly positive $f_{\phi}\,(\sim 5\,{\rm km\,s^{-1}\,kpc^{-1}})$ despite having large negative $\langle L_z^{\rm b} \rangle$ (and also $\normLz$).
  
In this section, we investigate central Splash stars in a sub-sample selected from our bar sample using $ f_{\phi} < 0 $ which comprises about $38\%$ of all stars with $\normLz < 0$. The left column of Fig.~\ref{fig: fig7} shows the fraction of stars with $\normLz <0$ (blue), $f_{\phi}<0$ (dashed green), and those associated with the retrograde ${\rm x_4}$ orbits (solid green) as a function of metallicity. The fraction of stars with $f_{\phi} < 0$ closely matches that of the ${\rm x_4}$ orbits, remaining nearly constant from the highest metallicities down to [Fe/H] $\sim -0.3$, before increasing from about $5\%$ at [Fe/H] $\sim-0.3$ to about $20\%$ at $\sim -1$ \citep[see also][]{Prudil2025}. Over the same metallicity range, the fraction of stars with $\normLz < 0$ rises from about $20\%$ to about $45\%$.

Following the same procedure as in Section~\ref{sec: 34} (Fig.~\ref{fig: fig5}), a five-component Gaussian fitting is performed to the MDF of stars with $f_{\phi}<0$ in our bar sample. The primary component (weight $\sim60\%$) of this Gaussian fit is shown as the dashed green curve in the right column of Fig.~\ref{fig: fig7}. The blue curve -- same as the primary component shown in Fig.~\ref{fig: fig5} -- shows the MDF of central Splash stars based on $\normLz<0$. Both MDFs agree very well, showing the prevalence of central Splash stars in the sub-sample with $f_{\phi}<0$.

In summary, bar region stars with $f_{\phi}<0$ in the metallicity range $-1<{\rm [Fe/H]}<-0.3$ are also part of the central Splash population, and they are dominated by the retrograde ${\rm x_4}$ orbits. Further, we note that the retrograde ${\rm x}_4$ orbits are generally confined primarily to the bulge \citep{Valluri2016unified, Pandey2025}.

\subsection{Ensemble dynamical properties of central Splash stars} \label{sec: 36}

\subsubsection{Rotational properties} \label{sec: 361}
\begin{figure}
	\centering
	\includegraphics[]{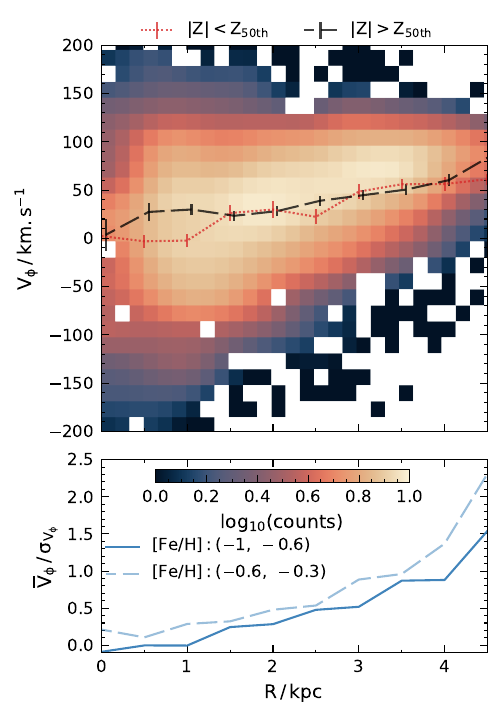}
	\caption{Top: Star count distribution of the central Splash stars in the plane of inertial-frame rotational velocity $V_{\phi}$ versus cylindrical radius $R$. The red and black lines show the running means with $R$ in the two intervals below and above the median vertical height distribution ($Z_{\rm 50th}\sim1.2$ kpc). Compared to the red curve, the black curve is shifted rightwards by 0.05 kpc for clarity. Bottom: The ratio of the mean observed Galactocentric rotational velocity $\overline{V}_{\phi}$ to the corresponding velocity dispersion $\sigma_{V_{\phi}}$ as a function of $R$ for central Splash stars in two metallicity intervals as indicated.}
	\label{fig: fig8}
\end{figure}

\begin{figure}
	\centering
    \vspace{12pt}
	\includegraphics[]{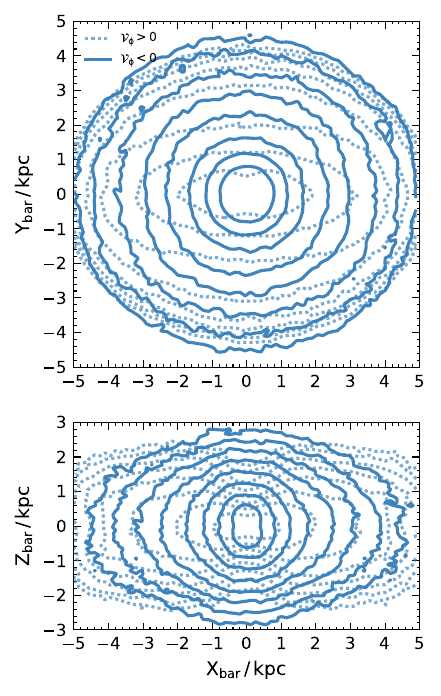}
	\caption{Orbital density distributions in the bar-frame for central Splash stars ($\normLz < 0$\,{\rm and}\,$-1 < {\rm [Fe/H]} < -0.3$; solid blue) and stars with $\normLz > 0$ in the same metallicity range (dotted blue), shown in face-on and edge-on views. Iso-density contours span $3.75$ to $2$ dex in steps of $0.25$ dex from inner to outer regions.}
	\label{fig: fig9}
\end{figure}

Fig.~\ref{fig: fig8} shows the rotational properties of central Splash stars ($\normLz<0$ and $-1<{\rm [Fe/H]}<-0.3$) using their observed Galactocentric inertial-frame rotational velocities. The top panel shows the star counts distribution in the $V_{\phi}-R$ plane. The distribution is nearly symmetrical around $V_{\phi}\sim0$ with velocity dispersion $\sigma_{V_{\phi}}\sim100\,{\rm km\,s^{-1}}$ in the central $\sim1.5$ kpc region. With increasing $R$, the distribution gradually shifts to higher $V_{\phi}$ and lower velocity dispersion, especially near the boundary of the bar sample ($R_{\rm apo}\leq 5$ kpc).

Also shown are the mean rotational velocity $\overline{V}_{\phi}(R)$ profiles for central Splash stars in two intervals of vertical height, above (dashed black) and below (dotted red) the median of the $|Z|$ distribution. Despite the slow rotation compared to bar populations \citep[see][]{Pandey2025}, the $\overline{V}_{\phi}(R)$ profile is nearly independent of vertical height for $R\gtrsim1$ kpc, implying that central Splash stars follow a cylindrical rotation pattern from the bulge to the outer bar. 

The bottom panel of Fig.~\ref{fig: fig8} shows the ratio $\overline{V}_{\phi}/\sigma_{V_{\phi}}$ as a function of $R$ for central Splash stars in two metallicity intervals as indicated. In both metallicity intervals, $\overline{V}_{\phi}/\sigma_{V_{\phi}}$ increases monotonically with $R$, accelerating as expected near the sample boundary. Additionally, at all radii $\overline{V}_{\phi}/\sigma_{V_{\phi}}$ is larger in the higher metallicity interval. 

\subsubsection{Orbital density and radial distributions} \label{sec: 362}
The face-on and edge-on orbital density distributions of central Splash stars (solid blue) are compared in Fig.~\ref{fig: fig9} with those of stars with $\normLz > 0$ in the same metallicity range (dotted blue). The orbital density distribution of stars with $\normLz > 0$ shows a clear bar in the face-on view and a pronounced peanut shape in the edge-on view. These structures are much more prominent than in the orbital density distributions for all stars \citep[][]{Khoperskov2024III, Pandey2025}. Outside the bar along the minor axis and towards corotation near the major axis the density distribution becomes oval-shaped; thus $\normLz>0$ stars are not all {\it bar-supporting}. Referring to their angular momentum we will call them {\it bar-following}.

In contrast, the orbital density distribution of central Splash stars is nearly axisymmetric in the face-on view and round to elliptical in the edge-on view, leading to an overall spheroidal density, quite different from the bar-following orbits. Central Splash stars neither follow the bar nor support its density.

\begin{figure}
	\centering
	\includegraphics[]{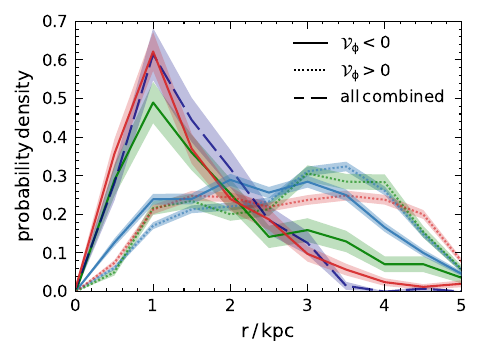}
	\caption{Galactocentric spherical distance distributions of high-[Mg/Fe] ($>0.2$) stars with $-1<$ [Fe/H] $<-0.3$ (blue) and [Fe/H] $>-0.3$ (green), low-[Mg/Fe] ($<0.15$) stars with [Fe/H] $<0.4$ (red) and [Fe/H] $>0.45$ (dashed navy). Solid, dotted, and dashed curves represent stars with $\normLz<0$, $\normLz>0$, and both combined.}
	\label{fig: fig10}
\end{figure}

Fig.~\ref{fig: fig10} shows Galactocentric spherical distance distributions of the stellar samples in Fig.~\ref{fig: fig9}, together with $\normLz<0$ and $\normLz>0$ samples of high-[Mg/Fe] ($>0.2$) stars with [Fe/H] $>-0.3$ (green), low-[Mg/Fe] ($<0.15$) stars with [Fe/H] $<0.4$ (red), and all stars with [Fe/H] $>0.45$ combined (navy). The distribution of central Splash stars is nearly uniform across the inner bulge and outer bar, and is very similar to their positive $\normLz$ bar-following counterparts.

In contrast, the radial distribution of the metal-rich ([Fe/H] $>-0.3$) high-[Mg/Fe] stars with $\normLz<0$ is strongly centrally concentrated to the bulge; while that of the $\normLz>0$ counterparts is radially extended to the outer bar. A similar striking difference in the radial distributions is seen between the retrograde and prograde $\normLz$ parts of low-[Mg/Fe] stars with [Fe/H] $<0.4$. Finally, the radial distribution of all extremely metal-rich ([Fe/H] $>0.45$) stars combined is strongly centrally concentrated, consistent with the recently discovered `knot' component \citep{Rix2024EMRknot, Khoperskov2025disc}. The central concentration of these populations in Fig.~\ref{fig: fig10} is consistent with their predominantly small apocentre distances $r_{\rm apo}\sim1-3$ kpc (middle-left panel of Fig.~\ref{fig: fig4}).

Overall, these results show that all $\normLz>0$ stellar samples have extended radial distributions, including those in the SplashMR. Among the $\normLz<0$ samples, central Splash stars are unique in showing a similarly extended radial distribution, whereas all other $\normLz<0$ samples are strongly concentrated toward the bulge.

\begin{figure}
	\centering
	\includegraphics[]{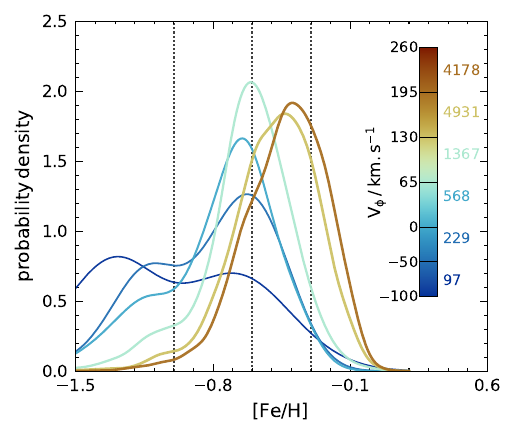}
	\caption{MDFs of \snhd\ high-[Mg/Fe] ($>0.2$) stars with $R_{\rm apo} > 5$ kpc from our base sample, shown in different observed $V_{\phi}$ intervals. Colours denote the lower velocity-bin edges, with the number of stars in each bin indicated. Dotted lines mark ${\rm [Fe/H]}=-1$, $-0.6$, and $-0.3$.}
	\label{fig: fig11}
\end{figure}

\subsection{Fraction of central Splash stars} \label{sec: 37}
Central Splash stars with $\normLz < 0$ and $-1<{\rm [Fe/H]}<-0.3$ account for $37\%$ of all high-[Mg/Fe] ($>0.2$) stars in our bar sample within this metallicity interval. They are relatively more numerous in the bulge ($45\%$, $R<2.5$ kpc) than in the outer bar ($27\%$, $R>2.5$ kpc)\footnote{We estimated that accounting for APOGEE photometric selection effects has only minor effect on the central Splash fraction, reducing it by $\sim 4\%$ for the similar bar sample of \citet{Pandey2025}, which lacks some additional high-latitude fields compared to the present sample and where we have corrected the weights for individual stars.}.

These central Splash fractions are substantially higher than those typically reported for Splash stars in the \snhd\ ($\sim5-10\%$), where Splash stars are identified using more halo-like dynamical selection criteria \citep[e.g.][]{Belokurov2020, Kisku2025}. Therefore we first estimate the effect of the different defining criteria. 

Our criterion $\normLz<0$ for the dynamically hot central Splash stars would also include disc stars that were dynamically heated rather than splashed through the GSE merger before being rearranged by the bar. We can assess this effect by looking at the MDF of \snhd\ high-[Mg/Fe] ($>0.2$) stars in bins of observed inertial-frame rotational velocity. Fig.~\ref{fig: fig11} shows that the MDF of our APOGEE base sample for $R_{\rm apo}> 5$ kpc evolves smoothly from a metal-rich distribution peaking at ${\rm [Fe/H]}\sim-0.4$ for high $V_{\phi}$, to a Splash-dominated distribution peaking at ${\rm [Fe/H]}\sim-0.6$) for $V_{\phi}<130\,{\rm km\,s^{-1}}$ (compare Fig.~\ref{fig: fig5} and \citealt{Myeong2022}). Still these low-$V_{\phi}$ stars are kinematically consistent with the hot thick disc \citep{Amarante2020heatedDisc, Binney2023}. Among \snhd\ high-[Mg/Fe] stars in our base sample within the range $-1<{\rm [Fe/H]}<-0.3$, those with $V_{\phi}<130\,{\rm km\,s^{-1}}$, i.e., Splash stars including heated disc stars, constitute a fraction of $\sim20\%$.    

A possible systematic effect could come from stars with $\normLz<0$ generated through the bar and buckling instabilities, as suggested by their presence at [Fe/H] $\simeq -0.1$ ($\sim 15\%$ fraction from Fig.~\ref{fig: fig7}), where the Galactic bulge component must have originated predominantly via secular processes \citep[see, e.g.,][]{Khoperskov2024III}, and by the model analysis in \cite{Prudil2025}. These stars would mostly be found in the bulge (Fig.~\ref{fig: fig10}). Thus the remaining $\sim 60\%$ bar-following stars in the SplashMR lead to an estimated $\sim 9\%$ contamination from secular processes to the central Splash fraction, which is thus reduced to $\sim 28\%$, albeit with considerable uncertainty. 

Several results indicate that stars with $\normLz<0$ and $-1 < {\rm [Fe/H]} < -0.3$ predominantly belong to the central Splash population. First, the 84th percentile of the apocentre distances of $\normLz<0$ stars changes rapidly through [Fe/H] $\sim -0.3$, transitioning from radially extended orbits in the SplashMR to centrally concentrated orbits at [Fe/H] $>-0.3$ (bottom-left panel in Fig.~\ref{fig: fig4}). Consistent with this, the Galactocentric radial distribution is nearly uniform from the inner bulge to the outer bar in the SplashMR, but is concentrated toward the inner bulge at higher metallicities (see Fig.~\ref{fig: fig10}). Second, the MDF of the central Splash indeed accounts for most $\normLz<0$ stars in the metallicity range $-1 < {\rm [Fe/H]} < -0.3$ (Fig.~\ref{fig: fig5}).

Overall, it seems possible (though uncertain) that the Splash fraction exhibits an intrinsic radial gradient, increasing moderately towards the Galactic centre. Using APOGEE DR17 data, \cite{Kisku2025} found that the Splash fraction in the \snhd\ increases slightly with decreasing $R$. Zoom-in cosmological simulations on the other hand predict a steep central concentration of the Splash population in Milky Way-like galaxies with GSE-like merger histories \citep[][]{Grand2020, Orkney_Laporte_2026}. 

\section{Discussion}\label{sec: 4}

\subsection{The central Splash population} \label{sec: 41}
The Splash population in the \snhd\ is characterized by stars with low and even retrograde $V_\phi$ velocities, high orbital eccentricities and large $z_{\rm max}$, and it is primarily found in the metallicity range (SplashMR) $-1 < {\rm [Fe/H} < -0.3$  \citep[][]{Bonaca2017, DiMatteo2019splash, Belokurov2020, Myeong2022, Chandra2024}. In the inner Galaxy, a Splash-like population is indicated by the shift of the $V_\phi$ distribution to lower velocities and the mild increase in the $V_R$ and $V_\phi$ velocity dispersions with decreasing [Fe/H] in the same metallicity range \citep[Fig.~\ref{fig: fig1}; see also][]{Lian2020milky}. However, this population is blurred with the bar population due to the non-axisymmetric gravitational potential of the Galactic bar.

We have employed and successfully tested a normalised, time-averaged bar-frame angular momentum parameter $\normLz$ (equation~\ref{eq: eq2}), to separate stars on bar-following orbits more robustly from stars on orbits that are retrograde or dynamically hot with respect to the rotating bar. Time-averaging in a realistic barred potential \citep{Portail_2017} smoothes out local variations along the orbit -- such as transient angular momentum reversals --  and normalisation by the average radial size of the orbit removes the main variations between the bulge and outer bar.

In $\normLz$-[Fe/H] diagrams we can display the properties of multiple stellar populations simultaneously, which we have done in this paper using a bar stellar sample selected from the APOGEE DR17 survey \citep{Abdurro2022apogeeDR17}, and summing over the entire bar region (Fig.~\ref{fig: fig4}). We find that in the SplashMR, the $\normLz$ distribution of the bar region stars has a strong extension to non-bar-following orbits ($\normLz < 0$), characterised by high median orbital eccentricity and $z_{\rm max}$, increased radial velocity variance, and a radial
extent similar to the bar stars (see also Fig.~\ref{fig: fig10}).

At the same time, little if any difference is seen in the [Mg/Fe] and [Al/Fe] abundance ratio distributions between stars on the main bar-supporting ${\rm x_1}$ orbit family and those on $\normLz < 0$ orbits in the SplashMR. This strongly suggests that both sets of stars share a common formation history. Because the ${\rm x_1}$ orbits in this metallicity range are believed to have originated in the high-$\alpha$ disc as it went through bar and boxy/peanut bulge formation \citep{DiMatteo2016discOrigin, Debattista2017fractionation, Fragkoudi2018disc, Pandey2025}, we infer that also the Splash-like population originated in the high-$\alpha$ disc. This is further supported by the smooth transition between the main bar component and the $\normLz < 0$ population in all dynamical and chemical quantities shown in Fig.~\ref{fig: fig4}.

Furthermore, the MDF of the $\normLz < 0$ population in the SplashMR is in very good agreement with the MDF of the \snhd\ Splash stars in our selected APOGEE base sample, using criteria from \cite{Belokurov2020} and \cite{Myeong2022} (Fig.~\ref{fig: fig5}). The [Mg/Fe] and [Al/Fe] distributions of the  bar and \snhd\ samples agree within errors in their median and dispersions as well (Fig.~\ref{fig: fig6}). Additionally, the Galactocentric radial and vertical velocity dispersions of the Splash-like stars ($\narsim 100$ and $\narsim 80\,{\rm km\,s^{-1}}$, respectively) are surprisingly consistent with the $108\pm19$ and $79\pm21\,{\rm km\,s^{-1}}$ measured for the Splash in the \snhd\ \citep{Belokurov2020}, strengthening the dynamical link between the two populations further. Lastly, the smooth kinematic transition between the high-$\alpha$ disc and Splash population in the \snhd\ is mirrored in Fig.~\ref{fig: fig4} for the bar and $\normLz$-populations.

We conclude that bar region stars with $\normLz < 0$ and $-1 < {\rm [Fe/H]} < -0.3$ represent the inner Galaxy counterpart of the \snhd\ Splash population, which we refer to as the {\it central Splash}. We estimate the fraction of central Splash stars relative to high-$\alpha$ stars in the SplashMR to be $\sim 30\%$, somewhat higher than a similar MDF-based estimate of the fraction of (kicked out and heated) Splash stars in the \snhd\ (Section~\ref{sec: 37}).

\subsection{Origin of the central Splash} \label{sec: 42}

The similarity between the chemo-dynamical properties of the central Splash and the \snhd\ Splash populations implies their common origin in a wide-spread event. The best candidate for such an event is the early merger of the Milky Way with the GSE -- based on the synchronous truncation of the derived age distributions for Splash stars in the \snhd\ and stars in the GSE, estimated at around $10$ Gyr ago by \cite{Gallart2019, Belokurov2020}.

\begin{figure*}
	\centering
	\includegraphics[]{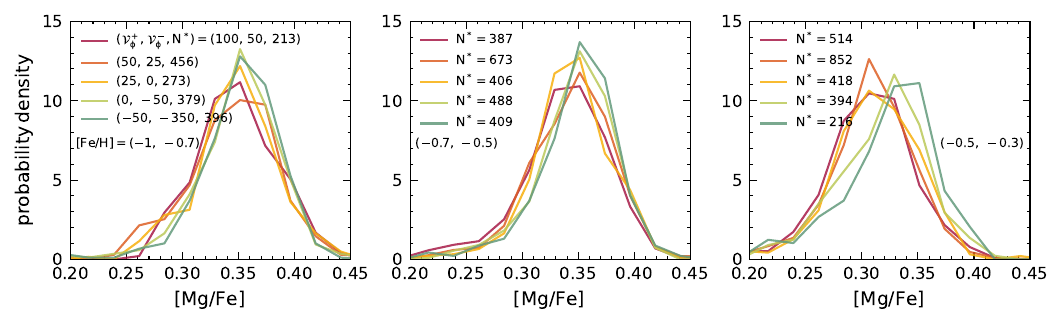}
	\caption{[Mg/Fe]-distributions in three metallicity intervals as specified on the panels. In each panel, different colours show intervals in $\normLz$ where the legend for each line gives the upper and lower bounds of the $\normLz$ interval, and the number of stars. From left to right the distributions for bar-following, $\normLz>0$ stars shift towards lower [Mg/Fe], while those for central Splash, $\normLz<0$ stars develop asymmetry in the rightmost panel.}
	\label{fig: fig12}
\end{figure*}

State-of-the-art cosmological zoom-in simulations of Milky Way analogues confirm the link between early gas-rich merger events and Splash-like populations \citep[][]{Belokurov2020, Grand2020, Renaud2021, Orkney_Laporte_2026}. GSE-like mergers heat the Milky Way proto-disc and scatter a fraction of disc stars onto low angular momentum halo-like orbits. A similarly 'seamless' transition between the Splash and the thick disk is predicted in simulation $V_{\phi}$-metallicity and $V_{\phi}$-age diagrams as is observed in the \snhd, and implied by the correspondingly seamless $\normLz$-metallicity maps for the central Splash. In several Auriga zoom-in models, the end of the merger was found to approximately coincide with both the truncation of the star formation in the GSE progenitor and the age of the youngest early disc stars scattered into the Splash \citep{Belokurov2020}. 

In a recent paper, \citet{Orkney_Laporte_2026} find typical mass fractions of splashed disc stars of $20\%-30\%$, with some high values at $75\%$, in a range of minor merger Auriga simulations (mass ratio $<$ 1~:~4), using their definition of Splash stars which includes retrograde and strongly heated prograde stars. These are compatible with our estimates in Section~\ref{sec: 37} above. Comparing with Milky Way constraints these authors argued that the GSE merger in the Milky Way was likely a minor merger with mass ratio of $\sim 1~:~7$. They also estimated a look-back time at age $11$ Gyr for the starburst induced by the first pericentric passage, corresponding to the formation of a sequence of metal-rich globular clusters. 

Alternative scenarios for the origin of Splash-like stars have also been proposed. With a hydrodynamic simulation of an isolated galaxy developing massive star-forming clumps during its early evolution \citep{Clarke2019clumps} and leading to the formation of both a low-$\alpha$ disc and a hot ($\sigma_{V_R}\narsim90\,{\rm km\,s^{-1}}$) high-$\alpha$ disc, \cite{Amarante2020splash} found that clump scattering can lead to a low-$V_{\phi}$ tail to $V_{\phi} \lesssim 0$ in the RotVD of the high-$\alpha$ disc stars.

The Splash population in the \snhd\ however appears as an additional component in the overall ROTVD, and it extends to $V_{\phi} \sim -130\,{\rm km\,s^{-1}}$ \citep{Belokurov2020}, well beyond what clump scattering produces. Also, the vertical velocity dispersion of Splash stars in both the bar region and the \snhd, $\sigma_{V_z} \sim 80\,{\rm km\,s^{-1}}$, is significantly higher than $\sigma_{V_z}= 50-60\,{\rm km\,s^{-1}}$ obtained  in \cite{Amarante2020splash} for the clump-driven Splash-like stars. This suggests that the possibly clumpy nature of the star forming early disc may not on its own be able to explain the full Splash velocity distribution.

More recently, \cite{Buder2026II} investigated the origin of Splash-like stars in a high-resolution NIHAO-UHD cosmological zoom-in simulation \citep{Buder2026I} of the last major merger of a Milky Way analogue. In this simulation, the merger follows a highly circular trajectory (circularity $\sim0.9$), causing little dynamical heating.  \cite{Buder2026II} found that proto-disc stars were born on low angular momentum orbits, and that the sickle-like transition between Splash-like and thick disc stars is seen in the metallicity interval corresponding to the spin-up. The Splash-like stars in this simulation do not reach as negative $V_{\phi}$ as in the \snhd, and the overall RotVD in this  interval is positively skewed, unlike for the \snhd\ \citep{Belokurov2022Dawn}, and the $\normLz$ distribution of the central Splash (Fig.~\ref{fig: fig4}). However, the oldest Splash and thick disc stars in this interval have identical [Mg/Mn]-[Al/Fe] distributions, reminiscent of Fig.~\ref{fig: fig6} above.

Overall, observational properties of the \snhd\ and central Splash, together with insights from cosmological zoom-in simulations, strongly favour their formation through the GSE merger which also explains the highly anisotropic, accreted inner halo of the Milky Way.

\subsection{The predicted starburst}\label{sec: 43}
The GSE merger simulations predict that a centrally concentrated starburst forming a thick disc is triggered by the first pericentre interaction \citep{Grand2020}. In minor mergers this dominates the Splash by a factor typically $2-4$ in mass \citep{Orkney_Laporte_2026}. In both $\sim$ 1~:~7 Auriga models analysed in their paper the Splash and starburst have broad MDFs, with the starburst MDF centered approximately on the upper half of the Splash MDF and reaching to slightly higher metallicities. The starburst has narrow age and [Mg/Fe] distributions (at lower mean [Mg/Fe] than the Splash), thus the burst is not strong enough to boost [Mg/Fe]), and a fraction of these stars can have Splash-like orbits. Evidence for these burst stars in the Milky Way has been elusive.

The decreasing central Splash fraction with metallicity in the APOGEE bar sample (Fig.~\ref{fig: fig7}) may suggest star formation at the higher metallicities in the Splash range. 
Fig.~\ref{fig: fig12} revisits the [Mg/Fe] distribution in three [Fe/H] bins in the SplashMR, in intervals of the $\normLz$ parameter, where large negative $\normLz$ signify the most retrograde average angular momenta relative to the bar (central Splash), and positive values the bar-following orbits. We find
that in all three metallicity bins, highly splashed cSplash stars peak at the highest [Mg/Fe]$\simeq0.34-0.36$. The [Mg/Fe]-distributions of bar-following stars are symmetrically centered on a similarly high [Mg/Fe] in the low [Fe/H] interval, but show a relative shift of $\sim -0.04$ in the highest [Fe/H] interval. 
Simultaneously the splashed stars develop lower-[Mg/Fe] tails, suggesting an additional component matching the [Mg/Fe] distribution of the bar-following stars, that for the more mildly splashed stars can be seen in the number counts in Fig.~\ref{fig: fig4}. These [Mg/Fe] variations are small, of order the individual star measuring accuracy (Sect.~\ref{sec: 21}) but the profiles in Fig.~\ref{fig: fig12} are each based on a few hundred stars.

We interpret the relatively high [Mg/Fe] of the most strongly splashed stars, and the invariance of the peak of their [Mg/Fe]-distributions with [Fe/H], in the sense that these are the stars kicked out earliest from the prior disc in the strong interaction during first pericentre. 
The shift in Mg/Fe in the upper [Fe/H] bin then signifies star formation after this time (because the lower-[Mg/Fe] stars were not affected by the initial strong splash), at slightly lower Mg/Fe because of an increased contribution from SNIa. This star formation must have been significant, based on the subsample sizes given in the figure, and because these [Mg/Fe], [Fe/H] are on the highest density ridge in the [Mg/Fe[-[Fe/H] distribution for the inner Galaxy \citep[e.g., Fig.~2 in][]{Pandey2025}. The occurrence of mildly splashed, newly formed stars in the highest metallicity interval could be due to weaker energy input during the brief final stages of the merger if the star formation took place in this period \cite[see][]{Orkney_Laporte_2026}, or due to the star formation occurring in a thick, dynamically hot gas disc.

To estimage the time at which this star formation occurred requires stellar ages. Several recent studies have been able to determine accurate ages for thick stars in the \snhd\ for RGB stars with astrometric constraints \citep{Miglio2021}, subgiant \citep{Xiang2022} and main sequence turn-off stars \citep{Sahlholdt2022}, and with Starhorse isochrone fitting \citep{Queiroz2023}. Overall, these studies obtained ages $\sim\!11\pm1.5$ Gyr.  The most accurate ages have been reported from the CMDGal program fitting stellar population models to CMDs for kinematically selected nearby thick disc samples \citep{Gallart2024, Gallart2025, FernandezAlvar2025}, which result in age determinations with observational and systematic errors of $\pm0.5$ Gyr.  From these results the main component of the nearby thick disc was formed $11$ Gr ago at [Fe/H]$\simeq -0.5$, with two smaller star formation peaks at ages $12$ Gyr and just before $10$ Gyr. 

An earlier study with a different age determination method found no radial age gradient across the high-$\alpha$ disc for $R=1-9$ kpc \citep{Imig2023tale}. If in the absence of sufficiently accurate age determinations in the bulge we can use thick disc ages in the \snhd, the formation of the bulk of stars with $-0.6\narlt {\rm [Fe/H]} \narlt -0.2$ ($-0.4\narlt {\rm [Z/H]} \narlt 0.$) would be placed \citep{FernandezAlvar2025} at $11\pm0.5$ Gyr ago with a tail to $\simeq 10$ Gyr at [Fe/H] around -0.45, dropping sharply thereafter. This is around and after the inferred time for GSE's first pericentre, suggesting that the strongest star formation in the Galactic high-[Mg/Fe] disc may have been triggered by the merging GSE satellite. However, more research is required to discriminate whether this star formation started before or only after the splashing, and how burst-like it was. 

\subsection{A classical bulge?}\label{sec: 44}

Could the central Splash population after all be the long-sought classical bulge (ClB) in the Milky Way? We argue that two of its main properties are not compatible with this possibility:

1) As illustrated in Sect.~\ref{sec: 34}, the MDF and the [Mg/Fe] and [Al/Fe] ratios of the central Splash are very similar to those of the Splash (in situ halo) in the \snhd. From this we infer that the central Splash originated from a very similar (disc) stellar population as the \snhd\ Splash. The \snhd\ Splash stars in turn were inferred to have been kicked out of the early Galactic disc by the strong perturbations during the merger with the GSE satellite -- based on their chemodynamics and the coincidence of the youngest ages seen in the Splash and GSE populations \citep{Bonaca2017, DiMatteo2019splash, Gallart2019, Belokurov2020}.

2) The central Splash population confined to the bar region extends to similar radii as the ($\normLz >0$) bar-following stars ($>4$kpc, Fig.\ref{fig: fig10}), and to vertical heights $\sim 2$ kpc, much larger than expected for a low-mass ClB. In fact, the persistence of low and retrograde rotation stars between the bar region and the \snhd\ \citep{Liao2024insights} makes it likely that the central Splash transitions smoothly into the \snhd.

ClBs are dense, centrally concentrated stellar systems found mostly in early-type spirals \citep{Kormendy2011, Fisher2016}. They continue the parameter correlations of elliptical galaxies towards lower luminosities and effective radii and are believed to have formed in early mergers before the discs of their host galaxies assembled \citep[see][]{Kormendy2011}.

The central Splash has typical metallicities [Fe/H] $\simeq -0.6\pm0.2$, at the limits of the observed samples of low-mass compact systems \citep{Gallazzi2005, Peng2015, LiTS2018}. We estimate a central Splash mass of $\sim 2-3\times 10^9\,M_\odot$ based on the fraction of such stars in [Fe/H] $=[-0.8,-0.5]$ (Sections~\ref{sec: 33}, \ref{sec: 37}) and the corresponding mass given by \cite{Khoperskov2024III}. ClB with such low mass have effective radii $\sim 100$ pc \citep{Erwin2015}, much smaller than both the present radial and vertical extent of the central Splash.

Galaxies with composite bulges containing more than one bulge type are described in, e.g., \cite{Erwin2015, Erwin2021}.  Only a few systems, including M31 \citep{BlanaDiaz2018}, clearly show signatures of both a ClB and a B/P bulge, due to difficulties of recognizing B/P bulges in barred galaxies at intermediate inclinations. In these systems the ClB is regularly found at the centre of the B/P bulge, contrary to the central Splash population which extends significantly outside the bulge into the outer bar region.

From both arguments combined we infer that the central Splash is not a separate ClB but that it formed from the high-[Mg/Fe] disc. Thus we come to a different conclusion from \cite{Nepal2025}. These authors separated an inner Galaxy sample into bar, disc, and a component of mostly metal-poor, high velocity dispersion, high eccentricity stars whose properties overlap with the central Splash obtained here, and which they interpreted as a pressure-supported spheroidal bulge pre-dating the thick disc.

Is there then any room left for a ClB? Returning to the $\normLz$-metallicity maps for the in situ populations in Fig.~\ref{fig: fig4}, we see no significant structure with $\normLz<0$ at low metallicities and high [Mg/Fe] other than the protogalactic
and central Splash populations. For the metal-poor protogalactic population \cite{Rix2022poor} estimate a small dust-corrected mass of $\gtrsim \times10^8 M_\odot$ and a Gaussian spatial extent of $\sigma_{R_{\rm GC}}\sim 2.7$ kpc. With the Splash linked to the high-[Mg/Fe]-disc and the GSE merger, this implies then that no significant ClB can exist in the inner Milky Way.  

This result strengthens previous results finding that a ClB could only represent a minor fraction of the total Milky Way bulge mass, based on (i) the observed cylindrical rotation \citep{Shen2010}, (ii) combined structural and kinematic constraints \citep{DiMatteo2014mapping}, (iii) dynamical models for the formation of the b/p bulge from combined thin and thick discs that were able to reproduce most of the structure and kinematics seen in early bulge surveys \citep{DiMatteo2015milky, Debattista2017fractionation, Fragkoudi2018disc}, and (iv) the model-independent result that the intrinsic $V_{\phi}$ rotation of most high-[Mg/Fe] bulge/bar stars is as fast as for the low-[Mg/Fe] bar stars; therefore the bulk of the high-[Mg/Fe] bulge must have a disc origin \citep{Pandey2025}.

Previously, a strong argument for a ClB in the Galaxy has been the very old ages of stars in the bulge ($\ge 10$ Gyr) obtained from modelling of colour-magnitude diagrams \citep[CMDs,][]{Clarkson2008, Brown2010, Renzini2018, Zoccali2026}. However, recently several studies have found similarly old ages for thick stars in the \snhd; see Section~\ref{sec: 43}. Based on CMD fitting of kinematically selected nearby thick disc samples \citep{FernandezAlvar2025}, the main component of the thick disc was formed $\sim11$ Gr ago at [Fe/H]$\simeq -0.5$, with two smaller star formation peaks at ages $12$ Gyr and just before $10$ Gyr. Thus the local thick disc appear to be as old as the old bulge populations. The high-[Mg/Fe] disc in the inner Galaxy would be expected to be at least as old. Thus the old bulge stars are consistent with being old (high-[Mg/Fe]) disc stars upscattered into the bulge by the Splash and by the later bar and buckling instabilities. 

In summary, our analysis of the central Splash leads to the conclusion that the Milky Way does not have an old classical bulge. The old bulge is the part of the high-[Mg/Fe] thick disk which is now in the bulge.

\subsection{Centrally concentrated super-metal-rich stars}\label{sec: 45}
The second stellar population in the bar region containing a noteworthy fraction of stars with $\normLz<0$ (i.e., negative bar-frame angular momenta) are the stars with super-solar metallicities ([Fe/H] $\gtrsim 0.3$; Figs.~\ref{fig: fig4}, \ref{fig: fig5}), which we discuss briefly here.  The most metal-rich of these stars are mostly confined to the bulge, with apocentre distances $r_{\rm apo} \sim 2$ kpc, and have large orbital eccentricities (Figs.~\ref{fig: fig4}, \ref{fig: fig10}). 

These properties agree with those of the extremely metal-rich stars in a similar APOGEE DR17 sample investigated by \cite{Khoperskov2025disc} which they found to be confined within $1$–$2$ kpc, move on radially biased orbits, and have an orbital density distribution with spheroidal morphology \citep[][]{Khoperskov2024III}. This population likely also corresponds to the central 'knot' identified previously by \cite{Rix2024EMRknot} in a large stellar sample with low-resolution Gaia XP spectra. These stars with ${\rm [M/H]_{XP}} \gtrsim 0.5$ display a round on-sky density distribution confined to the central $10^{\circ}$, corresponding to a projected distance of $\sim1.5$ kpc around the Galactic centre.

\cite{Khoperskov2025disc} speculated that the centrally confined extremely metal-rich star could have originated from the restart of star formation in the central regions following bar quenching, using up the remaining gas. The monotonic increase in the mean metallicity of low-$\alpha$ stars towards small radii \citep[][]{Eilers2022, Imig2023tale} indicated that no fresh gas was accreted into the central region. Deep within the Galactic potential well, the high gas pressure and efficient retention of metals released by newly formed stars could promote rapid chemical enrichment, producing extremely metal-rich stars that remain concentrated near the Galactic centre.

\section{Conclusions}\label{sec: 5}
We analysed the chemo-dynamical properties of in-situ stars in the Galactic bar region, specifically investigating the inner Galaxy counterpart of the \snhd\ Splash stars (in-situ halo). We selected a sample of in-situ stars with precise element abundances (Fe, Mg, and Al) from APOGEE DR17 spectroscopy, astrometry from Gaia EDR3, and distances from the AstroNN catalogue. For orbit parameters, we integrated orbits for all stars in a realistic gravitational potential with a rotating bar (with pattern speed $\Omega_{\rm b}=37.5\,{\rm km\,s^{-1}\,kpc^{-1}}$), from the suite of Made-to-Measure dynamical models developed by \cite{Portail_2017}, noting that results do not change in an alternate model with slightly lower pattern speed and differing in the shape of the potential. We excluded disc or halo interlopers with $R_{\rm apo}>5\,{\rm kpc}$ to focus on stars confined to the bar region.

\vspace{5pt}
\noindent Our main conclusions are:
\begin{itemize} 
    \item[(i)] The time-averaged bar-frame rotational angular momentum $\langle L_z^{\rm b} \rangle$ correlates with orbit types in the $f_R-f_{\phi}$ orbital frequency plane, and allows us to define and separate bar-following from non-bar-following orbits. Normalising $\langle L_z^{\rm b} \rangle$ by the mean orbital radius $\langle R \rangle$ yields a largely radius-independent quantity $\normLz$, which is well-suited for chemo-dynamical analysis of extended, intermixed stellar populations in the bar region.
\vspace{5pt}
    \item[(ii)] Bar region stars with $\normLz<0$ and $-1\narlt {\rm [Fe/H]} \narlt -0.3$ have chemical properties (MDF, [Mg/Fe], [Al/Fe]) that closely link them to \snhd\ Splash stars, and dynamical properties consistent with them. Their chemical properties also connect them closely to the high-[Mg/Fe] bar-following component with which they join smoothly in the $\normLz$-[Fe/H] plane. This provides compelling evidence that they formed in the early disc and represent the inner-Galaxy counterpart of the \snhd\ Splash population, which we refer to as the {\it central Splash}.
\vspace{5pt}
    \item[(iii)] We estimate that central Splash stars constitute $\sim 30\%$ of stars in the SplashMR ($-1 \narlt {\rm [Fe/H]} \narlt -0.3$) in our high-[Mg/Fe] ($>0.2$) bar sample. In the corresponding \snhd\ sample, stars with $-1 \narlt {\rm [Fe/H]} \narlt -0.3$ and $V_{\phi}<130\,{\rm km\,s^{-1}}$ for which the MDF is Splash-dominated account for about $25\%$ of high-[Mg/Fe] stars (splashed and strongly heated) in the same metallicity range, quite comparable to the fraction in the bar region. 
\vspace{5pt}
    \item[(iv)] The chemical similarity of the central Splash and \snhd\ Splash populations implies their common origin in a wide-spread event. Based on the smooth transition to the bar-following (\snhd\ thick disc) stars in the $\normLz$ ($V_{\phi}$) distributions in the SplashMR, the synchronous truncation of the age distributions of \snhd\ Splash and GSE stars., and the presence of the radially anisotropic accreted inner halo, the best candidate for this event is the Milky Way's early merger with the GSE satellite. 
\vspace{5pt}
    \item[(v)] Extremely metal-rich stars ([Fe/H]$\gtrsim0.4$) are centrally concentrated within $2$ kpc of the Galactic centre, consistent with the previously discovered `knot' component. This is the second population in the bar region containing a substantial fraction of non-bar-following stars, but at weaker $\normLz<0$ than the central Splash.  
\vspace{5pt}
    \item[(vi)]  The chemical similarity with the \snhd\ Splash, and its large radial extent similar to the bar-following stars, do not support the interpretation that the central Splash is a classical bulge such as observed in external galaxies. The $\normLz$-[Fe/H] in-situ stellar population maps do not contain a substructure with $\normLz<0$ at low [Fe/H] and high [Mg/Fe] other than the central Splash and the protogalactic stars. Therefore no significant classical bulge can exist in the Milky Way - the old Galactic bulge is consistent with the fraction of the old thick disc which has become part of the bulge through the splashing and the later bar formation.
\end{itemize}

We also found signatures of star formation after the initial splashing, in the form of a slight downward shift of the [Mg/Fe]-distributions for bar-following stars relative to central Splash stars in the upper part of the SplashMR. Comparison with thick disc ages and estimates for the time of splashing suggests that the strongest star formation in the Galactic high-[Mg/Fe] disc could have been triggered by the GSE merger. This connection and the nature of this star formation is an interesting subject for further studies with upcoming surveys.

\section*{Acknowledgements}  
\noindent Funded by the Deutsche Forschungsgemeinschaft (DFG, German Research Foundation) under Germany's Excellence Strategy - EXC-2094 - 390783311; Excellence Cluster ORIGINS, Boltzmannstr. 2, D-85748 Garching, Germany.
This work has used data from the Sloan Digital Sky Survey IV (SDSS, \url{https://www.sdss4.org}), funded by the Alfred P. Sloan Foundation, the U.S. Department of Energy Office of Science, and the Participating Institutions. SDSS acknowledges support and resources from the Center for High-Performance Computing at the University of Utah.
This work has also made use of data from the European Space Agency (ESA) mission Gaia (\url{https://www.cosmos.esa.int/gaia}), processed by the Gaia Data Processing and Analysis Consortium (DPAC, \url{https://www.cosmos.esa.int/web/gaia/dpac/consortium}). Funding for DPAC has been provided by national institutions, in particular the institutions participating in the Gaia Multilateral Agreement.

\section*{Data Availability}
The data underlying this paper will be shared on reasonable request to the authors.

\bibliographystyle{mnras}
\bibliography{references}

\appendix

\section{}

The figures presented here clarify several specific aspects of the discussion in the main text of Section~\ref{sec: 3}.

\begin{figure*}[!b]
	\centering
	\includegraphics[scale=1.175]{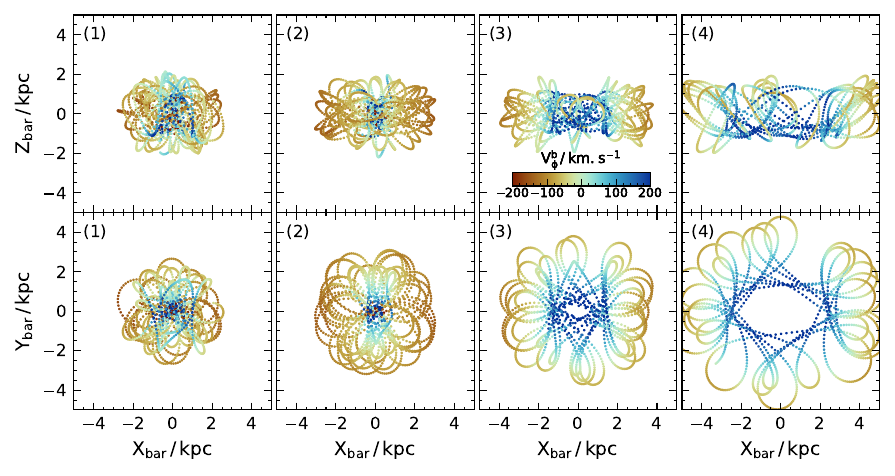}
    \caption{Representative orbits with $\normLz<0$ and $f_{\phi}>0$, selected based on their $f_R$ and $f_{\phi}$ frequencies as indicated in Fig.~\ref{fig: fig2}. Each column shows two projections of one orbit in bar frame coordinates. Points along the orbit are colour-coded by the bar-frame rotational velocity $V_{\phi}^{\rm b}$, where the colour bar is common to all panels. Positive $V_{\phi}^{\rm b}$ corresponds to the sense of Galactic rotation, which is clockwise in the $X_{\rm bar}-Y_{\rm bar}$ plane.}
	\label{fig: figA1}
\end{figure*}

\begin{figure*}
	\centering
	\includegraphics[]{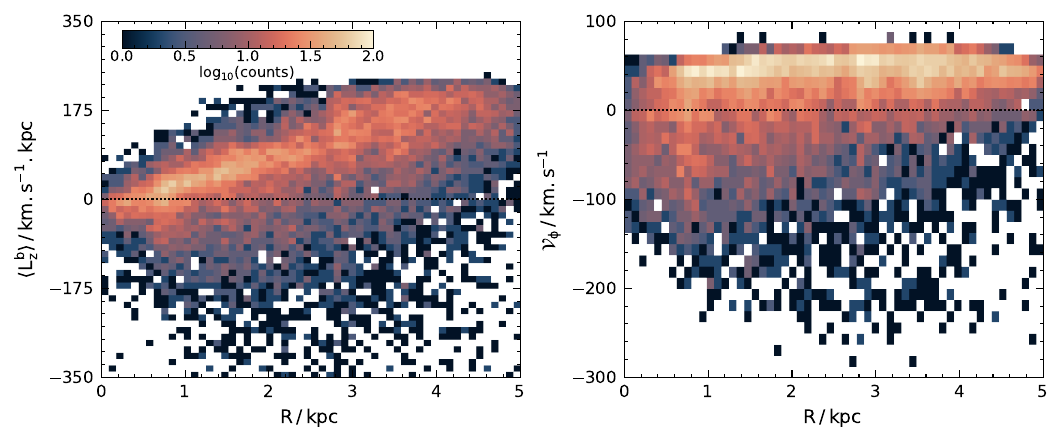}
	\caption{The left and right panels show the dependence of the time-averaged $\langle L_z^{\rm b} \rangle$ and normalised time-averaged $\normLz$ bar-frame rotational angular momentum on the observed Galactocentric radius $R$ for all stars in our bar sample. The colour scale, which is common to both panels, represents the number of stars per pixel.}
	\label{fig: figA2}
\end{figure*}

\begin{figure}
	\centering
	\includegraphics[]{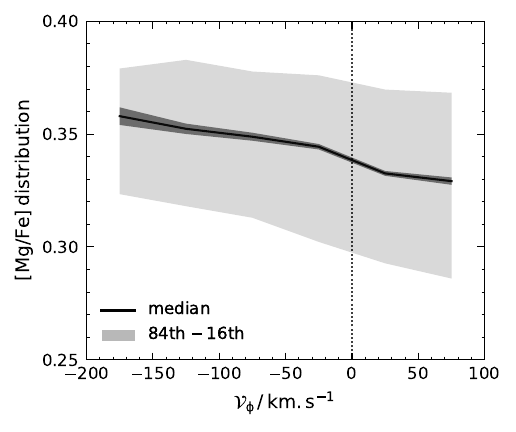}
	\caption{The [Mg/Fe] distribution of stars in the SplashMR, $-1 < {\rm [Fe/H] < -0.3}$, as a function of $\normLz$ for stars in our bar sample. The black solid line and the width of the dark grey band shows the median [Mg/Fe] and its bootstrapped uncertainties. The width of the light grey band represents the difference of the $84$th and $16$th percentiles of the distribution.}
	\label{fig: figA3}
\end{figure}

Fig.~\ref{fig: figA1} shows the orbital structures of representative orbits with $\normLz<0$ and $f_{\phi}>0$, selected according to their $f_R$ and $f_{\phi}$ values in Fig.~\ref{fig: fig2}. These orbits mainly differ in $f_R$, which decreases from orbit-(1) to orbit-(4). Orbit-(1) is nearly round and three-dimensional, and remains mostly confined to the bulge. As $f_R$ decreases, the orbits reach larger radii and become less vertically extended. They also develop a multi-foil structure in their outer parts with a central hole. In all cases, these orbits have a large positive bar-frame rotational velocity near pericentre and a smaller but negative velocity near apocentre. This gives rise to the sign mismatch between $\normLz$ and $f_{\phi}$.

Fig.~\ref{fig: figA2} compares the radial dependence of the bar-frame $\langle L_z^{\rm b} \rangle$ and $\normLz$ for stars in our bar sample. The distribution of stars in the $\langle L_z^{\rm b} \rangle$–$R$ plane in the left panel exhibits a strong positive correlation with the Galactocentric radius $R$. In contrast, the $\normLz$–$R$ distribution in the right panel shows little or no systematic dependence on $R$, demonstrating that the normalisation by the mean orbital radius $\langle R \rangle$ used in equation~(\ref{eq: eq2}) effectively removes the radial trend in $\langle L_z^{\rm b} \rangle$.

Fig.~\ref{fig: figA3} shows the [Mg/Fe] distribution of bar sample stars in the SplashMR as a function of $\normLz$. The median [Mg/Fe] increases from $0.33$ to $0.36$ dex as $\normLz$ decreases from $\sim75$ to $\sim -180,{\rm km,s^{-1}}$. However, this slight increase towards negative $\normLz$ remains within the $\sim1\sigma$ range around the median, indicating that the [Mg/Fe] distribution is nearly homogeneous across the full range of $\normLz$ for stars in the SplashMR.

\bsp
\label{lastpage}
\end{document}